\documentclass[%
reprint,
superscriptaddress,
 amsmath,amssymb,
aps,
pra,
]{revtex4-2}

\newcommand{\ket}[1]{\left|#1\right\rangle}

\usepackage{pifont} 
\usepackage{graphicx}
\usepackage{dcolumn}
\usepackage{bm}
\usepackage{hyperref}
\usepackage[mathlines]{lineno}
\usepackage{xcolor}
\usepackage{ bbold }

\begin{document}

\preprint{APS/123-QED}

\title{Characterizing Physical Error Contributions of Quantum Gates}

\author{Miha Papi\v c}
\email{miha.papic@iqm.tech}
\affiliation{IQM Quantum Computers, Georg-Brauchle-Ring 23-25, 80992 Munich, Germany}
\affiliation{Department of Physics and Arnold Sommerfeld Center for Theoretical Physics, Ludwig-Maximilians-Universität München, Theresienstr. 37, 80333 Munich, Germany}

\author{Adrian Auer}
\affiliation{IQM Quantum Computers, Georg-Brauchle-Ring 23-25, 80992 Munich, Germany}

\author{Inés de Vega}
\affiliation{IQM Quantum Computers, Georg-Brauchle-Ring 23-25, 80992 Munich, Germany}
\affiliation{Department of Physics and Arnold Sommerfeld Center for Theoretical Physics, Ludwig-Maximilians-Universität München, Theresienstr. 37, 80333 Munich, Germany}

\date{\today}

\begin{abstract}
Large-scale quantum computation requires a reliable assessment of the main sources of error in the implemented quantum gates. To this aim, we provide a learning-based framework that extracts the contribution of each physical error source to the infidelity of a series of gates, together with an uncertainty estimate for every contribution. Crucially, the uncertainties provided by the underlying Gaussian Process Regression serve as a diagnostic of the assumed noise model itself: if the physical model does not fully explain the experimental observations, the framework signals this instead of returning a falsely precise error budget. We demonstrate the proposed budget learning procedure on a pair of qubits in a transmon array architecture, where we compute the error budgets of the native single- and two-qubit gates. For this purpose, we have also developed an advanced noise model based on the first-principles understanding of the error processes in current superconducting processors, including complex (e.g. non-Markovian) error sources. While the single-qubit budgets are reconstructed with high confidence, the same diagnostic reveals that even this detailed noise model does not fully capture the error dynamics of a state-of-the-art two-qubit gate --- quantitative evidence that uncertainty-aware error budgeting is necessary for the characterization of current quantum hardware.

\end{abstract}

\maketitle


\section{\label{sec:intro} Introduction}

In recent years, there has been a steady advance in the scale and quality of quantum computing architectures, although noise and imperfections remain a vital issue holding back quantum advantage. An accurate diagnosis of the physical origin of these errors is essential for improving quantum hardware, optimizing calibration procedures, and developing noise-aware error mitigation \cite{cai2022quantum} and error correction \cite{Krinner_2022, google_qec_2022, google_qec_2024}. Such diagnosis requires not only estimating gate fidelities but also identifying which physical mechanisms contribute to the observed errors.

The noise present in quantum circuits is typically modeled by appending noise channels after each gate in the algorithm \cite{cheng_2021, Dahlhauser_2021, Donatella_2021, piskor2025}. These noise channels can be reconstructed experimentally, or constructed from a predefined model together with experimentally measured parameters. This approach relies on several implicit assumptions, namely that the noise is temporally and spatially uncorrelated, trace-preserving, and sufficiently small in the operator norm, i.e. that treating the noise as acting after an ideal zero-duration gate is accurate to first order in the noise strength \cite{Li_2014}. These assumptions are known to be violated in realistic hardware, particularly in superconducting qubits where leakage and temporal correlations have been extensively reported \cite{Motzoi_2009,Braumuller_2020,Sung_2021,Krinner_2022,Marxer_2022, miao_2022}.

This has motivated characterization methods beyond the Markovian assumptions. Techniques such as Gate Set Tomography (GST) \cite{Nielsen2021gatesettomography} and Randomized Benchmarking (RB) \cite{Knill_2007} have been extended to capture certain non-Markovian effects \cite{White_2020,Figueroa-Romero_2021,figueroaromero2023operational,White_2022, Li_2024}, and phenomenological models have been proposed to describe complex dynamics in superconducting devices \cite{zhang_2022,tripathi_2023,agarwal2023modelling}; however, these methods primarily characterize the effective behaviour of the hardware rather than directly identifying the underlying physical error mechanisms.

To address these limitations, we formulate physical error budgeting as a simulation-based inverse inference problem.We train Gaussian Process Regression (GPR) surrogate models on simulated characterization experiments generated from a first-principles physical model of the hardware. Once trained, the surrogate directly infers the contribution of each physical error mechanism from experimentally accessible observables while providing uncertainty estimates for the inferred budgets. Importantly, these uncertainties are not merely confidence intervals but also indicate whether the assumed physical model adequately explains the experimental observations.

The main contributions of this work are threefold:
\begin{enumerate}
\item We formulate physical error budgeting as a simulation-based inverse inference problem.
\item We introduce a Gaussian Process Regression surrogate model that enables efficient inference, with a predictive uncertainty that serves as a quantitative diagnostic of the completeness of the assumed noise model.
\item By constructing a highly detailed noise model of the device we experimentally demonstrate the methodology by reconstructing the physical error budgets of native single- and two-qubit gates on a superconducting quantum processor, where the diagnostic detects that even this noise model does not fully capture the two-qubit gate dynamics.
\end{enumerate}

\section{\label{sec:error_budgeting_theory} Error Budget Reconstruction}

\begin{figure*}[t]
\centering
\includegraphics[width=.9\textwidth]{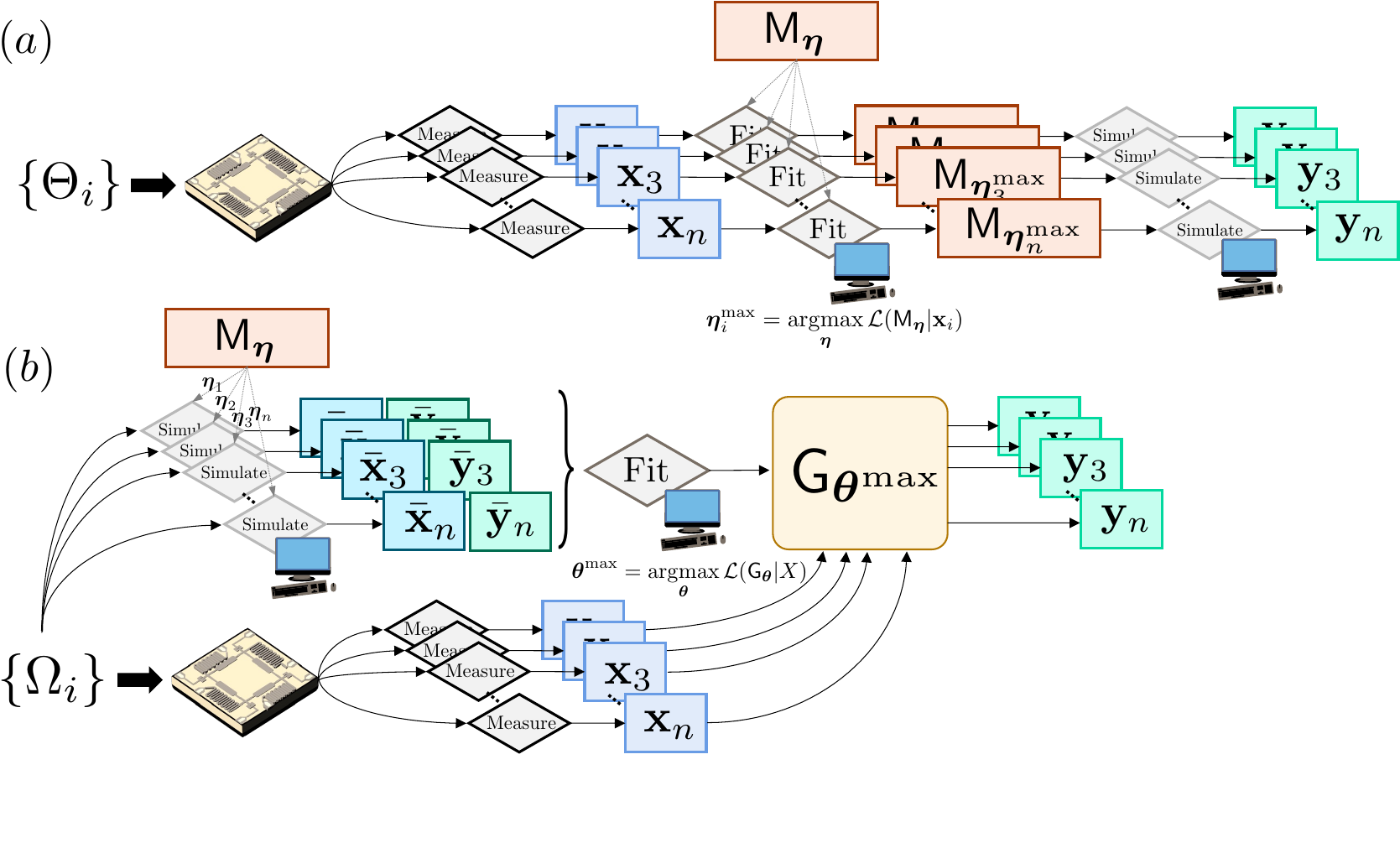}
\caption{\label{fig:budget_schematic} General schematic of error budget reconstruction of a number of quantum gates. Panel (a) depicts the standard approach, while the proposed alternative is illustrated in panel (b). The proposed approach is based on computing a surrogate function $\mathsf{G}$ which maps experimental results to error budget contributions directly.
}
\end{figure*}

The extraction of a physical error budget of a quantum gate requires a \emph{dynamical} physical noise model, based on our understanding of the possible error mechanisms, as well as a means to characterize the parameters of the chosen model. A dynamical noise model is any model that can predict and simulate the desired quantum operation and therefore evaluate a performance measure for the error budgets. Due to its widespread adoption, we use the infidelity of a quantum gate as this performance measure throughout this work. A dynamical noise model $\mathsf{M}$ therefore maps a set of noise parameters $\mathbf{\eta}$ to a set of experimental measurements $\mathbf{x}$ or error contributions $\mathbf{y}$, and can be evaluated either numerically or analytically. The complete budgeting procedure typically comprises the following steps:
\begin{enumerate}
    \item Perform a set of characterization experiments $\{ \Omega_i \}$ and obtain a set of experimental observations $\{ \mathbf{x}_i \}$.
    \item Determine the parameters of the noise model $\mathbf{\eta}$ by fitting the observed data e.g. by using a maximum-likelihood estimation procedure, $\mathbf{\eta}_i^\mathrm{max} = \mathrm{argmax}_\mathbf{\eta} \mathcal{L}(\mathsf{M}_\eta|\mathbf{x}_i)$, where $\mathcal{L}$ is the likelihood function.
    \item Evaluate the noise model with the obtained parameters to extract the error budget contributions, $\mathsf{M}_{\mathbf{\eta}_i^\mathrm{max}} \rightarrow \mathbf{y}_i $.
    \item Repeat previous steps for all gates and error sources of interest.
\end{enumerate}
Note, however, that in the above workflow, the noise model and the parameter characterization are treated as separate steps, which can lead to untested assumptions and approximations. A verification step can be performed but is not typically done, so the error contributions are extracted from the noise model with the obtained parameters without further validation, which may lead to unreliable estimates.

To illustrate this, consider the infidelity contribution of a Markovian pure-dephasing $T_\phi$ decay on a single-qubit gate in a transmon architecture. The noise model $\mathsf{M}$ is characterized by performing a Ramsey experiment. The value of $T_\phi$ can then be extracted by fitting an exponential decay to the measured probabilities for different waiting times. The duration of the gate $t_g$ is a known parameter set by the experimentalist. Therefore $\mathbf{\eta} = [T_\phi, t_g]$ and the noise model $\mathsf{M}_{[T_\phi, t_g]}$ maps these parameters to the infidelity contribution of the $T_\phi$ decay as e.g. $\mathsf{M}_{[T_\phi, t_g]} \approx t_g / (6T_\phi)$ \cite{Papic_2024}. Nonetheless, in this procedure we have implicitly assumed that the $T_\phi$ times of a driven and idling qubit are identical, and this assumption is not explicitly tested. As we will show in the following, such an estimate is unreliable due to (a) non-Markovian contributions to the pure dephasing noise, which are not captured by the simple exponential decay model, and (b) the $T_\phi$ time of a driven qubit can be significantly different from that of an idling qubit.

This example illustrates how characterization experiments using different operations than those we are interested in budgeting may lead to untested approximations. This approach is illustrated schematically in Fig. \ref{fig:budget_schematic}(a). It also quickly becomes unpractical when we wish to characterize several gates and error sources in the presence of complex noise processes, such as non-Markovian noise and non-computational states: a single evaluation of the dynamics with a complex noise model can be computationally costly, and the optimization procedure to find the best parameters is typically non-convex, requiring a large number of evaluations. Furthermore, if we wish to characterize several gates, the characterization procedure must be run independently for each gate, introducing a significant classical computational overhead.

\section{\label{sec:gpr_for_budgets} Gaussian Processes for Error Budget Reconstruction}

To overcome this issue, without introducing the need for additional model verification experiments, we propose a different approach, illustrated in Fig. \ref{fig:budget_schematic}(b). Our proposal obtains the function $\mathsf{G}_\theta$, which maps experimental observations $\mathbf{x}$ to error budget contributions $\mathbf{y}$ directly. Here, $\theta$ denotes the set of parameters of the function $\mathsf{G}_\theta$. Assuming we have an accurate noise model $\mathsf{M}$ of the device, supervised learning techniques can compute $\mathsf{G}_\theta$ based on a large sample of simulated characterization experiments with known error budget contributions. The classical computational effort required to generate the training dataset is performed offline and only once for a given device model.

A supervised learning model for $\mathsf{G}_\theta$ in Fig. \ref{fig:budget_schematic}(b) must have the following properties: (a) it can model non-linear relationships between the experimental observations and the error budget contributions, (b) it can be trained with a small number of data points, since simulating the dynamics of a quantum system with complex noise processes can be computationally costly, and (c) it can provide uncertainty estimates for its predictions, to estimate the confidence of the prediction and, equally importantly, the consistency between the experimental observations and the assumed physical model. Large predictive uncertainties therefore directly indicate that relevant physical mechanisms are missing from the model $\mathsf{M}$.

A learning technique which satisfies all the above requirements is Gaussian Process Regression (GPR) \cite{rasmussen_gpr}. More information about Gaussian Process Regression is available in Appendix~\ref{app:GPR}.

The complete budgeting procedure illustrated in Fig.~\ref{fig:budget_schematic}(b) is therefore comprised of the following steps:
\begin{enumerate}
    \item Randomly select parameters of the noise model $\mathbf{\eta}_i$ from a predefined range of values, chosen based on prior knowledge of the device (see Appendix~\ref{app:exp_details} for the ranges employed in this work), and simulate the dynamics of the system with the noise model $\mathsf{M}_{\mathbf{\eta}_i}$ to obtain a set of simulated experiments $\{\Omega_i \}$ with outcomes $\bar{\mathbf{x}}_i$ and error budget contributions $\bar{\mathbf{y}}_i$.
    \item Train a GPR model $\mathsf{G}_\theta$ with the simulated data $\{ \bar{\mathbf{x}}_i, \bar{\mathbf{y}}_i \}$ by maximizing the log-likelihood of the model parameters $\theta$, where $\theta^\mathrm{max} = \mathrm{argmax}_\theta \mathcal{L}(\mathsf{G}_\theta|\{ \bar{\mathbf{x}}_i, \bar{\mathbf{y}}_i \} )$.
    \item Perform the set of experiments $\{ \Omega_i \}$ on real hardware and obtain a set of experimental observations $\{ \mathbf{x}_i \}$.
    \item Use the trained GPR model to predict the error budget contributions $\mathbf{y}_i$ and their corresponding uncertainties based on the experimental observations $\mathbf{x}_i$, $\mathsf{G}_{\theta^\mathrm{max}}[\mathbf{x}_i] \rightarrow \mathbf{y}_i$.
    \item Repeat the above steps for each error source of interest.
\end{enumerate}

\begin{figure}[t]
\centering
\includegraphics[width=.45\textwidth]{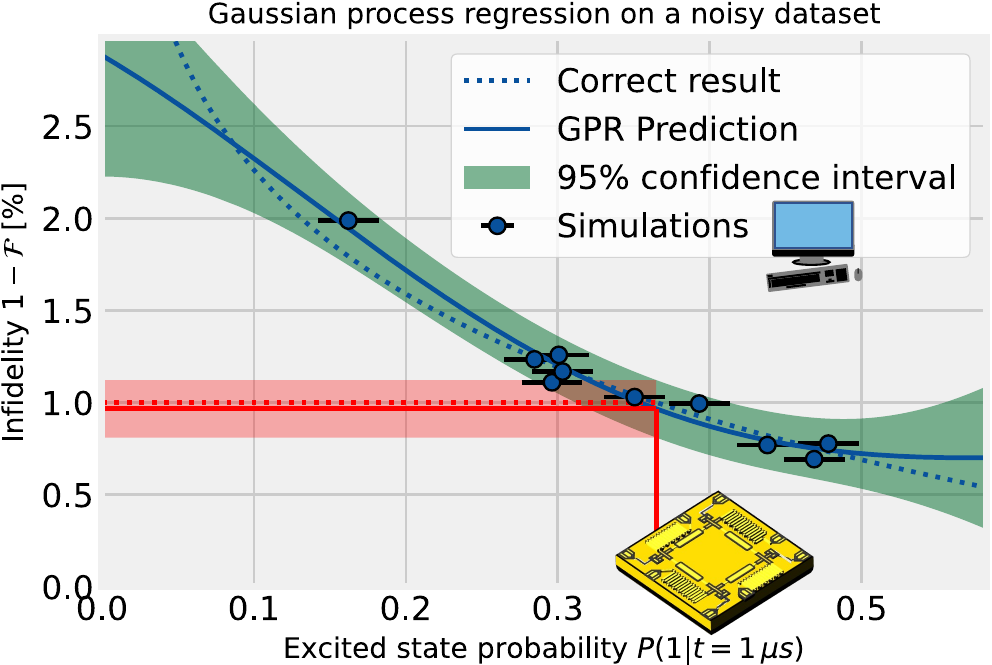}
\caption{\label{fig:gpr_demo}  
An illustrative example of a simplified scheme to measure the $T_1$ decay induced infidelity using a single input experiment and GPR. The input experiment used in this case consists of preparing a qubit in the excited state and waiting for 1 $\mu$s before measuring the excited state ($x$-axis). The data points correspond to a number of simulated qubit evolutions with different $T_1$ decay times, and different measurement errors indicated by the horizontal error bars. An interpolation of these points using GPR is plotted with the solid blue line, with the shaded area representing the uncertainty of the GPR prediction, which in this case is due to state preparation or measurement errors. As a reference, the exact relationship between the measured excited state probability and the infidelity of the idling operation ($y$-axis) is plotted with the dotted blue line. In general, this relationship is typically not known, due to the presence of different error sources, and numerical simulations are the only available probe. According to the proposed method, when an experimental result is obtained (vertical red solid line), the GPR prediction curve would allow to determine the corresponding infidelity (horizontal red solid line). Furthermore, an estimate of the accuracy of this prediction due to the GPR prediction is depicted with the shaded red area. The correct value is plotted with the horizontal dotted red line.
}
\end{figure}

To further illustrate this approach, consider the simple task of identifying the contribution of $T_1$ decay to the infidelity of the idling operation, as illustrated in Fig.~\ref{fig:gpr_demo}. For illustration purposes, we will use a single experiment to attempt to characterize the $T_1$. We initialize the qubit to the excited state, wait for a certain time, and then measure the probability of the qubit remaining in the excited state. By simulating this experiment for different values of $T_1$ together with different measurement errors, we can train a GPR model to predict the infidelity contribution of the $T_1$ decay based on the measured excited state probability at the chosen time. We then use the experimental result as an input to the GPR model to estimate the corresponding infidelity contribution and its uncertainty, as shown in Fig.~\ref{fig:gpr_demo}. From there it is also evident that if the experimental result is significantly different from the simulated ones, the GPR model will return a large uncertainty, giving us a clear signal that our noise model does not adequately explain the experiment. This behaviour follows directly from the structure of the GPR: the predictive variance grows with the distance, as measured by the optimized kernel, between the input and the training data (Appendix~\ref{app:GPR}). Even in this simple example, if the infidelity of the idling operation were instead dominated by a different error source, with a different relationship between the measured excited state probability and the infidelity, the GPR model would similarly return a large uncertainty for the experimental result.

The proposed framework should be viewed as a physics-informed inverse inference procedure. As with any inverse problem, the quality and uniqueness of the inferred error budgets depend on the information contained in the chosen characterization experiments together with the expressiveness of the underlying physical model.

Distinct physical error mechanisms may produce sufficiently similar experimental signatures that they cannot be uniquely distinguished from the available observations. If such ambiguity is present, it is fundamental and cannot be removed by the regression model itself. It can only be reduced by designing more informative characterization experiments or by incorporating additional prior physical knowledge. The predictive uncertainty produced by the Gaussian Process naturally reflects regions where this ambiguity is present. 

We also define the error budgets as corresponding to the isolated contribution of each physical error mechanism. While different error sources are approximately additive in the weak-noise regime \cite{abad2022,abad2023}, interactions between multiple simultaneous error mechanisms may generate higher-order contributions. Such effects could naturally be incorporated by extending the framework to infer higher-order error budgets involving multiple interacting error sources.

Finally, the scalability of the proposed methodology is limited primarily by the scalability and accuracy of the underlying physical simulator. The framework is therefore intended for the characterization of individual gates or small subsystems, rather than entire quantum processors, consistent with how calibration and characterization procedures are performed in practice. We note that the qubit-coupler-qubit system underlying the two-qubit gate demonstration in Sec.~\ref{sec:budgeting_demo} is the largest primitive that needs to be characterized in the considered architecture, and other superconducting gate schemes are unlikely to be significantly more complex.

\section{Demonstration of the Error Budgeting Procedure}\label{sec:budgeting_demo}

We now demonstrate the error budgeting procedure described in the previous section on both single and two-qubit gates in a transmon-based superconducting architecture with tunable couplers, as depicted in Fig.~\ref{fig:experiments}(a). To demonstrate the proposed inference methodology we employ a detailed first-principles noise model of a superconducting transmon processor. The simulations are used as the training data required by the surrogate model and include the dominant coherent, incoherent and temporally correlated error mechanisms expected in current devices.

To model the dynamics as accurately as possible, the simulations are performed at the pulse level \cite{agarwal2023modelling}. More information about the simulator is found in Appendix~\ref{app:simulations}.

The budgeting procedure was implemented on a 5-qubit superconducting quantum computer \cite{iqm_spark_2024}. The native gates on this platform are the $\sqrt{X}$ single-qubit and a controlled-Z (CZ) two-qubit gate, which together with virtual-Z rotations form a universal gate set. The $\sqrt{X}$-gate is implemented using a microwave Gaussian Derivative Removal by Adiabatic Gate (DRAG) pulse \cite{Motzoi_2009}, while the CZ is implemented diabatically with a tunable coupler that induces a Rabi oscillation between the levels $|11\rangle \leftrightarrow |02\rangle$ to obtain the conditional phase. Both gate implementations have been used in large-scale transmon-based experiments, and have demonstrated state-of-the-art performance on testing devices \cite{Marxer_2022, google_qec_2022, google_qec_2024, marxer_2026}.

The characterization circuits employed here were deliberately chosen to remain experimentally simple and avoid requiring pulse-level access. Consequently, the protocol is not guaranteed to be informationally complete for arbitrary physical error models. Rather, it is designed to distinguish the dominant error mechanisms expected in present-day superconducting processors while simplifying the characterization procedure.

\subsection{Single-Qubit Gate Noise Model}\label{subsec:SQG_noise}

\begin{table}[t]
\caption{List of single-qubit gate error sources included in the noise model. Whether the error source is Markovian is indicated in the second column, while the parameters used to model the error source are listed in the third column. An explanation of the model parameters is found in Appendix~\ref{app:sqg_noise_model}. The term ``Pulse'' indicates that the error contribution depends on the pulse parameters. The parameters below the horizontal line were included in the simulation, however their contributions to the budget are negligible, and therefore they are not included in the final budget estimates. They were nonetheless included in the simulation for completeness.
}
\label{tab:SQG_errors}
\begin{ruledtabular}
\begin{tabular}{ccc}
Error Source & Markovian & Parameters \\
\hline
Amplitude Damping ($T_1$) & \ding{51} & $T_1$  \\
Markovian Dephasing ($T_\phi$) & \ding{51} & $T_\phi$ \\
$1/f$ Noise & \ding{55} & $T_{1/f}$ \\
Over- and Under-rotation (I) & \ding{51} & $\epsilon_A$ \\
\hline
DRAG pulse amplitude error (Q) & \ding{51} & $\epsilon_\beta$ \\
Axis misalignment & \ding{51} & $\epsilon_\mathrm{phase}$ \\
Frequency detuning & \ding{55} & $\delta \omega$ \\
Leakage & \ding{55} & $\omega$, $\alpha$, Pulse
\end{tabular}
\end{ruledtabular}
\end{table}

A full list of all modelled single-qubit gate error mechanisms is available in Table~\ref{tab:SQG_errors}. The parameter ranges for all the listed error sources were chosen based on previous characterization experiments performed on the same device or with the same electronics setup. The parameter ranges for the Hamiltonian of the system (qubit frequency and anharmonicity) are obtained based on the predicted design values and the deviations observed in typical devices. Further information on the exact modelling of all the errors presented here is available in Appendix~\ref{app:sqg_noise_model}.

The characterization experiments used for the single-qubit budget reconstruction are visualized in Fig.~\ref{fig:experiments}(b), composed of applying various combinations of $\sqrt{X}$ gates in conjunction with noiseless virtual Z rotations. More specifically, the characterization experiments are constructed as follows:
\begin{enumerate}
    \item Apply 21 $\sqrt{X}$ or $\sqrt{X}^\dagger$ gates to the qubit. Note that $\sqrt{X}^\dagger$ can be implemented using one physical $\sqrt{X}$ and virtual Z rotations.
    \item Optionally apply either a 200 ns idling duration, or 21 more $\sqrt{X}$ or $\sqrt{X}^\dagger$ gates.
    \item Optionally apply either a 200 ns idling duration, or 21 more $\sqrt{X}$ or $\sqrt{X}^\dagger$ gates.
\end{enumerate}
Altogether, the characterization experiments employ between 21 and 63 physical operations, for a total of 26 characterization experiments. These circuits are designed to separate the error sources listed in Table~\ref{tab:SQG_errors}: repeating the gate amplifies the errors, with coherent contributions accumulating quadratically and incoherent ones linearly, while the optional idling periods distinguish the decoherence of a driven qubit from that of an idling one, in the spirit of the example discussed in Sec.~\ref{sec:error_budgeting_theory}. Moreover, depending on the chosen combination of $\sqrt{X}$ and $\sqrt{X}^\dagger$ gates, coherent rotation-angle errors either accumulate or echo away, further separating them from the incoherent contributions.

\subsection{Two-Qubit Gate Noise Model}\label{subsec:TQG_noise}

\begin{table*}[t]
\caption{List of two-qubit gate error sources included in the noise model. Whether the error source is Markovian is indicated in the second column, while the parameters used to model the error source are listed in the third column. The term ``Ham.'' is used to indicate that the hybridization of the system \cite{marxer_2026, Yuan_2020, Marxer_2022} affects the contribution of that error source significantly. The term ``Pulse'' indicates that the error contribution depends on the pulse parameters. An explanation of the model parameters is found in Appendix~\ref{app:tqg_noise_model}.}
\label{tab:TQG_errors}
\begin{ruledtabular}
\begin{tabular}{ccc}
Error Source & Markovian & Parameters \\
\hline
Qubit (1,2) Amplitude Damping ($T_1$) & \ding{51} & $T_1$, Pulse, Ham.  \\
Qubit (1,2) Markovian Dephasing ($T_\phi$) & \ding{51} & $T_\phi$, Pulse, Ham. \\
Qubit (1,2) $1/f$ Noise & \ding{55} & $T_{1/f}$, Pulse, Ham. \\
Coupler Amplitude Damping ($T_1$) & \ding{51} & $T_1$, Pulse, Ham.  \\
Coupler Markovian Dephasing ($T_\phi$) & \ding{51} & $T_\phi$, Pulse, Ham. \\
Coupler $1/f$ Noise & \ding{55} & $T_{1/f}$, $\omega_C^\mathrm{max}$, Pulse, Ham. \\
Control Pulse & \ding{51}/\ding{55} & $A$, $\delta A$, $\tau$, $\delta \tau$ \\
Qubit (1,2) Virtual Z Error & \ding{51} & $\Delta \Phi_Z$ \\
Flux tails & \ding{55} & $A_n, \, \tau_n$ \\
\end{tabular}
\end{ruledtabular}
\end{table*}

The full list of all modeled two-qubit gate error sources is available in Table~\ref{tab:TQG_errors}. The parameter ranges for all the listed error sources were chosen based on previous characterization experiments performed on the same device or with the same electronics setup. The parameter ranges for the Hamiltonian of the system (qubit and coupler frequencies and anharmonicities) are obtained based on the predicted design values and the deviations observed in typical devices. Further information on the exact modelling of all the errors presented here is available in Appendix~\ref{app:tqg_noise_model}.

The characterization experiments used for the CZ characterization are shown in Fig. \ref{fig:experiments}(c). The state preparation consists of preparing each of the qubits in the $(\ket{0} + \ket{1})/\sqrt{2}$, such that the two-qubit density matrix does not contain any zero entries. Afterwards, we apply the CZ gate 3, 5, and 7 times and perform two-qubit state tomography. These experiments are designed such that the errors of the CZ gate are amplified, while also containing enough information about the different error sources. For example, the rate at which the errors accumulate with the number of gates contains information about the coherence of the errors. All together, the total number of characterization experiments is 48. We additionally perform a correlation analysis based on the Chatterjee coefficient on the simulated data and use only the 20 most correlated input experiments in the budget characterization to reduce the experimental cost, while still maintaining a good accuracy of the budget estimates. We note that even though the state-tomography procedure in the presence of state-preparation and measurement errors cannot reconstruct the state exactly, this is not problematic for the budgeting procedure, as long as the same errors are modeled in the simulated data.

\subsection{State Preparation and Measurement Errors}\label{subsec:SPAM_errors}

State Preparation and Measurement (SPAM) errors do not contribute to the error budget of the gate, but are necessary to accurately model the experimental observations. An imperfect modelling of the SPAM errors will therefore lead to a mismatch between the experimental observations and the simulated ones, resulting in less reliable budget estimates. While error amplification in the input experiments helps mitigate this issue, an accurate model of the SPAM errors is still beneficial.

Based on previous experiments on superconducting qubits which were also confirmed to be present in the considered device, we model the state preparation errors as a residual thermal population with an effective temperature in the range of $40-60$ mK. The measurement errors are modelled with a classical confusion matrix, based on the previously observed readout errors on the same device, which are in the range of $1-3\%$ for the single-qubit case. More details about the SPAM error modelling are available in Appendix~\ref{app:state_preparation} and~\ref{app:measurement}.

\subsection{Experimental Results}\label{subsec:experimental_results}

\begin{figure*}[t]
\centering
\includegraphics[width=.95\textwidth]{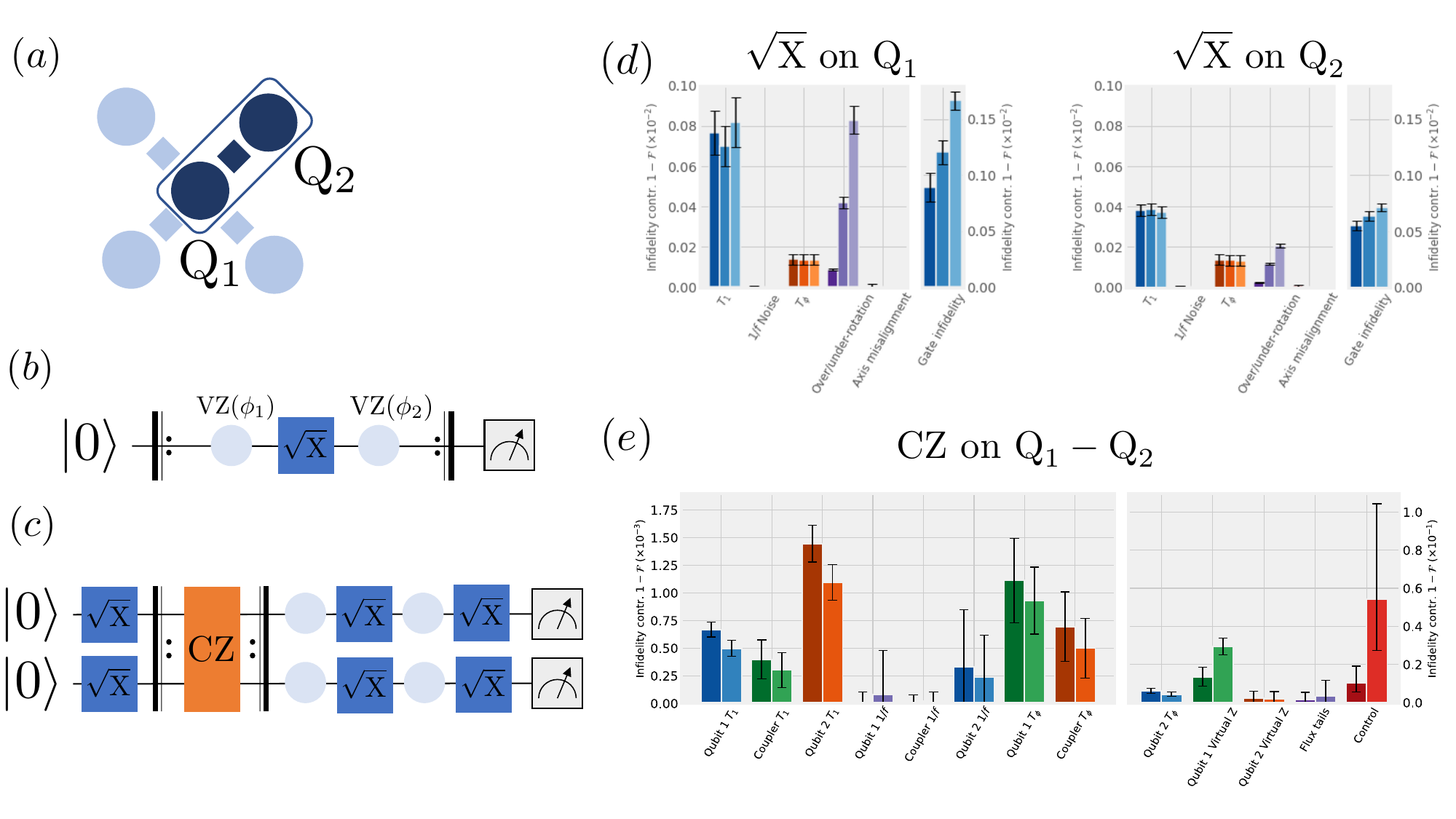}
\caption{\label{fig:experiments} Error budgets of the native single- and two-qubit gates of an IQM Spark device. (a) The topology of the device, with the highlighted qubit pair used in the demonstration. (b) A circuit representation of the single-qubit input experiments from which the budgets are computed. (c) A circuit representation of the two-qubit input experiments from which the budgets are computed. The circles represent virtual Z rotations. (d) The obtained per-gate error budgets for the single-qubit gate on both qubits, with the error contributions of each source plotted in different colors. The error contributions are plotted with their corresponding uncertainties, which are obtained from the GPR model. The different columns represent the error budget contributions to the infidelity of one, three and five $\sqrt{X}$ gates. (e) The obtained error budgets for the two-qubit CZ gate, with the error contributions of each source plotted in different colors. The error contributions are plotted with their corresponding uncertainties, which are obtained from the GPR model. The two columns represent the error budget contributions to the infidelity of one and three CZ gates.
}
\end{figure*}

The described noise model was then used to simulate the dynamics of the system for a large number of noise parameter sets, and the resulting data trained a GPR model for each error source. Approximately 400 samples were obtained for all the budgets. The trained GPR models were then used to predict the error budget contributions based on the experimental observations obtained from the input experiments shown in Fig. \ref{fig:experiments}(b) and (c). The obtained error budgets are shown in Fig. \ref{fig:experiments}(d) and (e) for the single- and two-qubit gates respectively. The error contributions are defined as the infidelity of the gate if only the specified error is present (see Appendix~\ref{app:gate_performance}), and are plotted with their corresponding uncertainties obtained from the GPR model. The different columns represent the per-gate error budget contributions for 1, 5 and 9 repetitions of the $\sqrt{X}$ gate, and 1 and 3 repetitions of the CZ gate. Further details about the employed parameters of the noise model as well as of the GPR model are available in Appendix~\ref{app:exp_details}.

For the two single-qubit gate errors in Fig. \ref{fig:experiments}(d) we observe confident predictions of the GPR model for the dominant error sources, which are the Markovian amplitude damping, dephasing and the over- and under-rotation errors. The high confidence of the predictions allows us to also predict the total per-gate infidelity (on the right). As expected, the per-gate contribution of incoherent errors is constant, while the coherent errors accumulate quadratically with the number of gates. The contribution of the non-Markovian errors in this case is significantly smaller, and while the exact value is obscured by other error sources, the GPR model is still able to upper limit the effect of these errors to a small contribution to the infidelity.

On the other hand, for two-qubit gates in Fig. \ref{fig:experiments}(e) we split the error sources based on the magnitude of their contribution to the infidelity. In this case, incoherent errors related to the amplitude damping, dephasing and non-Markovian $1/f$ noise are secondary error sources, while the dominant error sources are the control pulse errors and the virtual Z rotation errors. The contribution of the flux tails is also significant, but a confident estimate is not possible due to the other error sources present. Overall, the budget contributions in this case are generally less certain compared to the single-qubit case. This is the model diagnostic of the framework at work: even though the employed noise model is a highly detailed first-principles error model of a diabatic two-qubit gate, the reconstruction quantitatively detects that it does not fully capture the error dynamics of the device. A budgeting procedure without uncertainty quantification would instead have attributed the observed infidelity to the modelled error sources with false confidence. The results thus identify the dominant error sources, while the uncertainty estimates indicate which contributions are reliable and where the noise model requires further refinement.

\subsection{Cost Comparison}\label{subsec:cost_comparison}
We can also approximately estimate how the errors from Fig. \ref{fig:experiments} can be diagnosed without using the GPR method outlined in this paper. Ignoring non-Markovian effects, methods such as GST \cite{Nielsen2021gatesettomography} and variants of randomized benchmarking \cite{Knill_2007,Figueroa-Romero_2021} can obtain more accurate benchmarks, but at a much higher experimental cost. To make the comparison fair, we consider the error characterization procedures from Ref.~\cite{Sung_2021} as a baseline, since we can make the same assumptions about the error sources as in our GPR based procedure.

As a reminder, the GPR procedure described in this paper requires, for the two-qubit gate, three state tomography experiments, or 48 different circuits.

\begin{table}[h]
\caption{\label{tab:tqg_comparison_table}
The characterization experiments performed in Ref. \cite{Sung_2021} that are needed to obtain similar results to Fig. \ref{fig:experiments}, with the same assumptions about the noise. }
\begin{ruledtabular}
\begin{tabular}{cc}
Error Source  & Characterization experiment \\
\hline
Qubit 1 $T_1 $ &  $T_1$ measurement  \\
Qubit 2 $T_1 $ &   $T_1$ measurement   \\
Control &  State tomography + $f$-state readout \\
Flux tails & Several Ramsey exp. + SWAP to Coupler \\

\end{tabular}
\end{ruledtabular}

\end{table}

The experiments required to characterize the same error sources with comparable accuracy, but without the GPR method, are listed in Table \ref{tab:tqg_comparison_table}. To elaborate on the data in Table \ref{tab:tqg_comparison_table}, we assume a $T_1$ experiment with modest accuracy requires $\sim$5 circuits, while a single two-qubit state tomography requires 16 distinct circuits and the leakage contribution can be roughly estimated with a single circuit, provided $f$-state readout is available. For the Ramsey experiments needed to evaluate the flux-tail contribution, we assume at least $\sim 10$ further circuits are needed, as an accurate coupler frequency characterization requires several oscillations to be seen in the experimental data. Further assuming that the coupler frequency after applying a flux pulse is measured at $\sim$5 different delay times to estimate the flux-tail magnitude, this brings the \textit{minimal} total number of distinct circuits needed for the characterization to $\sim 75$, about a factor of $\sim 1.5$ more than the proposed GPR based procedure.

Additionally, the coupler transmon typically cannot be directly read out and does not have dedicated drive lines for implementing single-qubit gates. Probing the flux-tail effects on the coupler therefore requires first preparing an excitation in the computational transmon and then transferring it to the coupler with a SWAP-like operation, which needs additional experiments to tune up; the same applies to the coupler readout. We estimate that this additional calibration, assuming the computational transmon frequency is known from the calibration of single-qubit gates, requires $\sim 25$ more data points: to maximize the population transfer between the computational and coupler transmons, we must tune the coupler to the frequency of the qubit transmon and let the system interact for a time of approximately $1/g_{2c}$, so that one Rabi cycle is performed. If the frequency of the coupler and duration of the interaction are swept with 5 distinct values each, we arrive at $\sim 5\times 5 = 25$.

Together with the previous steps, this adds up to a total of 100 distinct circuits to run, resulting in a final advantage of more than a factor of $\sim 2$ for the GPR based method.

Not using the GPR method also has additional drawbacks:
\begin{enumerate}
    \item Once the noise parameters are experimentally obtained, a subsequent simulation of the system with them is needed to evaluate the actual infidelity contributions. Translating the measured uncertainties into fidelity bounds will demand even more simulation time. The time associated with the classical simulations depends on the complexity of the system and error sources, but if no analytical formulas are available, in our case it can take up to a couple of minutes, comparable to the time needed to run the experiments on the quantum hardware. In the GPR case, the classical simulation can be performed in parallel and beforehand.
    \item The GPR based procedure uses circuit outcomes as inputs and requires no pulse-level control of the system. Probing any error source affecting the coupler (flux tails, coupler $1/f$, etc.) will always demand pulse-level access due to the aforementioned SWAP-like operation.
\end{enumerate}

\section{Conclusions}\label{sec:conclusions}

In this work we have formulated physical error budgeting as a simulation-based inverse inference problem in which experimentally accessible observables are mapped directly to physical error contributions through surrogate models trained on first-principles simulations. The predictive uncertainty produced by the surrogate model should be interpreted not only as a confidence interval but also as a diagnostic tool indicating possible discrepancies between experiment and the assumed physical model. Such discrepancies may reveal previously neglected physical mechanisms and therefore naturally guide future improvements of the noise model. Although demonstrated here on a superconducting transmon processor, the methodology itself is independent of the underlying quantum computing platform and can in principle be applied to any architecture for which sufficiently accurate physical simulations are available, such as co-design chips \cite{renger_2026} or ion-trap systems \cite{Lotshaw_2022}. 

Using the developed noise model, we have demonstrated that current superconducting hardware is not necessarily limited purely by environmental effects such as $T_1$ decay, since control imperfections can be a significant source of error in an algorithm execution. This fact is often hidden when reporting gate fidelities as a performance metric, since the infidelity of a single gate can be dominated by incoherent errors, while the infidelity of a series of gates or an algorithm can be dominated by other error contributions, due to the different scaling of coherent errors. These findings warrant coherent error suppression techniques such as randomized compiling \cite{Hashim_rc_2021}, as well as more frequent calibration. Interestingly, we have also found that the non-Markovian nature of $1/f$ flux noise implies that we will often overestimate the effect of the pure dephasing decay on the gate infidelity, if the pure dephasing decay times obtained via Ramsey experiments are used in the estimation.

A central result of this work is that the uncertainty of the budget estimates is generally larger for the two-qubit gate compared to the single-qubit case: the framework itself detects that the employed noise model, while highly detailed, has not fully captured the dynamics of the errors. One possible reason could be additional, unmodelled pulse distortions, such as reflections in the drive lines. Since our model also takes into account the effects of state hybridization during gate operation \cite{marxer_2026, Yuan_2020, Marxer_2022}, the dynamics depend on the noise spectrum of the device at previously unprobed frequencies in the MHz regime, which can lead to a mismatch between the simulated and experimental observations. Additionally, the random nature of the frequency dependence of the amplitude damping processes in the device can never be modelled exactly, which would imply that a reliable error budget for the gate can only be obtained by measuring the transmon decay over a wide range of frequencies \cite{Bylander_2011, Burnett_2019}. Due to non-computational elements, such a characterization of the full system is challenging \cite{Marxer_2022, marxer_2026}. Our results therefore suggest that further improvements in the noise modelling are still needed to obtain more accurate budget estimates for diabatic two-qubit gates.

As discussed in Sec.~\ref{sec:gpr_for_budgets}, the reported contributions correspond to the isolated effect of each physical mechanism rather than to a strictly additive decomposition of the total gate infidelity; extending the framework to higher-order budgets involving interacting error sources is a natural direction for future work.

Furthermore, the characterization experiments employed here were defined by the requirement of ease of implementation: we restricted the circuits to native gates and standard measurements, avoiding pulse-level access, and then selected the most informative circuits within this class via a Chatterjee correlation analysis of the simulated data (Appendix~\ref{app:exp_details}). A natural extension of this analysis is a quantitative identifiability study: computing the sensitivity of the simulated observables with respect to each error-source parameter, e.g. in terms of the Fisher information, would determine which error sources a given set of circuits can distinguish in principle, and would provide a principled basis for the circuit selection. Combining such an analysis with Bayesian optimal experiment design would then allow characterization experiments to be selected adaptively so as to maximize the information gained about the underlying physical error mechanisms under the same implementability constraints \cite{Sarra_2023, ramoa2025}, improving the overall accuracy of the error budgets while reducing the total experimental overhead.

As the number of qubits increases in the future, the inference methodology itself remains unchanged, provided that the errors remain local. In the opposite case, more scalable, but less accurate noise models will need to be employed, such as a near-Clifford gate-based simulator.

\begin{acknowledgments}
We thank all of the employees at IQM for fruitful discussions, especially Kuan Yen Tan for contributing to the original idea, and also Alessio Calzona and Amin Hosseinkhani for revising the manuscript. We acknowledge support from the German Federal Ministry of Education and Research (BMBF) under Q-Exa (grant No. 13N16062) and QSolid (grant No. 13N16161).
\end{acknowledgments}
\appendix

\section{Gaussian Process Regression}\label{app:GPR}

In this section we briefly overview the basics of Gaussian Process Regressors as implemented in this work, starting by defining a Gaussian Process (GP) before moving on to the regression task.

A multivariate Gaussian probability distribution is defined as 
\begin{equation}
    \mathcal{N}(\mathbf{x},\pmb{\mu},\Sigma) = \frac{1}{(2\pi)^{D/2}|\Sigma|^{1/2}}\exp{\left[-\frac{1}{2}(\mathbf{x} - \pmb{\mu})^T\Sigma^{-1}(\mathbf{x} - \pmb{\mu}) \right]},
\end{equation}
which is completely specified by its mean vector $\pmb{\mu} \in \mathbb{R}^D$ and covariance matrix $\Sigma$. The off-diagonal elements of the covariance matrix $\Sigma$ determine the degree of correlation between the elements of the vector $\mathbf{x} = (x_1,x_2,...,x_D)^T$. If one were to draw random samples of $\mathbf{x}$ and plot them in pairs $\{(i,x_i), \,1 \leq i \leq D \}$: if $\Sigma$ is a diagonal matrix, each element of the vector is drawn independently from the others, and the resulting plot would display one realisation of uncorrelated white noise. On the other hand, if $\Sigma$ has large off-diagonal elements, each element of $\mathbf{x}$ is strongly correlated with the others, and the resulting plot reminds us of a smooth function. This intuitive example paves the way to understanding how to use Gaussian processes for supervised learning tasks.

Instead of individually specifying each element of the covariance matrix $\Sigma$, we will always parametrize it in terms of covariance functions $\left[ \Sigma \right]_{i,j} = k(\mathbf{x}_i,\mathbf{x}_j)$. Furthermore, if $k(\mathbf{x}_i,\mathbf{x}_j)$ depends only on the scalar product of the vectors $\mathbf{x}_i \cdot \mathbf{x}_j$, the function $k(\cdot,\cdot)$ is a kernel function. 

An important property of the above distribution is the so-called marginalization property, meaning that if we arbitrarily split a multivariate Gaussian variable vector into two $\mathbf{x} = (\mathbf{x}_1,\mathbf{x}_2)^T$, both subvectors are also Gaussian distributed, $\mathbf{x}_{1,2} \sim \mathcal{N}(\mathbf{x}_{1,2},\pmb{\mu}_{1,2},\Sigma_{1,2})$, with the relevant submatrix $\Sigma_{1,2}$ \cite{rasmussen_gpr}.

A Gaussian process is a type of stochastic process, defined as a collection of random variables, any finite number of which have a joint Gaussian distribution. It is therefore completely specified by its mean $\pmb{\mu}$ and covariance matrix $\Sigma$. A Gaussian process model therefore describes a probability distribution over possible functions that fit a set of points \cite{intuitive_guide_to_GPR}.

Focusing on the regression task, we have access to a number of (possibly noisy) input-output pairs $\{ (\mathbf{x}_i,y_i)\}$, where $y_i = f(\mathbf{x}_i) + \epsilon$ and $\epsilon$ is an independently and identically distributed Gaussian noise in the observations. Instead of trying to find a single function $f(\mathbf{x})$ that best fits the data, we try to find a probability distribution over the functions that fit the data, and make predictions based on the means of this distribution of functions. Since we have a supervised learning task, we need to incorporate our knowledge of the prior observations.

The regression function defined by Gaussian processes is therefore
\begin{equation}
    P(\mathbf{y} | X) = \mathcal{N}(\mathbf{f} | \pmb{\mu},\Sigma),
\end{equation}
where $\mathbf{y} = (y_1, y_2,\dots,y_D)$ is a vector of our observations, and $X = (\mathbf{x}_1, \mathbf{x}_2,\dots,\mathbf{x}_D)$ is the ordered matrix of the input vectors corresponding to each observation. We have still included the mean vector $\pmb{\mu}$ although the input data is typically normalized to zero mean. The covariance matrix $\left[ \Sigma \right]_{i,j} =\mathrm{cov}\left[f(\mathbf{x}_i), f(\mathbf{x}_j) \right] = k(\mathbf{x}_i, \mathbf{x}_j)$ has been parametrized in terms of a kernel function. A typical, and most frequently used, example of a kernel function is the so-called radial basis function (RBF) or squared exponential
\begin{equation}\label{eq:rbf_kernel}
    k_\mathrm{RBF}(\mathbf{x}_i,\mathbf{x}_j) = A\exp\left(-\frac{|\mathbf{x}_i-\mathbf{x}_j|^2}{2 l^2} \right).
\end{equation}
While the amplitude of the kernel $A$ determines the standard deviation of a single point sampled from the Gaussian process (the diagonal values of $\Sigma$), the length scale $l$ is what determines the cross-correlation, and therefore how continuous the functions we wish to approximate are. If $l$ is very small, each point of the Gaussian process is sampled almost independently from its preceding point, and the functions resemble white noise, while for very large $l$ the Gaussian process is more reminiscent of a continuous function.

Furthermore, instead of using a single RBF, a sum of such functions of the form
\begin{equation}\label{eq:kernel}
    k_{\Sigma} (\mathbf{x}_i,\mathbf{x}_j) = \sum_k A_k\exp\left(-\frac{|\mathbf{x}_i-\mathbf{x}_j|^2}{2 l_k^2} \right)
\end{equation}
is also a valid kernel function. Typically, we use 3 to 4 RBF kernels to achieve good performance. In a practical setting, it is often easy to see when increasing the number of kernels used is no longer necessary, as the optimized length scale $l_k$ and amplitude will both converge towards values close to zero, meaning that only white noise is being added to the model. 

In such a parametrization of the kernel function the parameters $\theta = \{ (A_k, l_k), k = 1,2,\dots \}$ are what determine the distribution over the functions and must therefore be optimized so that they best fit the observed data. This is done by maximizing the log marginal likelihood 
\begin{align}\label{eq:log_marg_likelihood}
    \log P(\mathbf{y} | X, \theta) &= -\frac{1}{2} \mathbf{y}^T (\Sigma + \sigma^2 \mathbb{1})^{-1} \mathbf{y} \nonumber\\
    &- \frac{1}{2}\log\mathrm{det}(\Sigma + \sigma^2 \mathbb{1})\\
    &- \frac{D}{2}\log(2\pi)\nonumber
\end{align} 
which is done with the limited memory version of the Broyden–Fletcher–Goldfarb–Shanno (L-BFGS) optimization algorithm \cite{scikit-learn}.

We account for the observations $\mathbf{y} = f(\mathbf{x}) + \epsilon$ being also inherently noisy, with Gaussian noise of mean zero and standard deviation $\sigma$. For simplicity we assume all the inputs have the same amount of noise $\sigma$ and we exclude heteroscedastic effects.

After the optimal kernel function parameters $\theta$ have been identified, we make predictions at new inputs $X_* = (\mathbf{x}_1^*,\mathbf{x}_2^*,\dots )$.

It can be shown that the distribution of the possible outputs $f(\mathbf{x}_*)$ is 
\begin{equation}
    P(f(\mathbf{x}_*) | X, X_*, \mathbf{y}) \sim \mathcal{N}(\Sigma_*^T \Sigma \mathbf{f},\Sigma_{**} - \Sigma_*^T \Sigma^{-1}\Sigma_* ).    
\end{equation}
We have used the following notation to denote the covariance matrices with different combinations of training and test (*) data
\begin{align}
&[\Sigma]_{ij} = k(\mathbf{x}_i,\mathbf{x}_j), \\
&[\Sigma_*]_{ij} = k(\mathbf{x}_i,\mathbf{x}_j^*), \\
&[\Sigma_{**}]_{ij} = k(\mathbf{x}^*_i,\mathbf{x}^*_j) .
\end{align}
From this distribution we can infer the mean and prediction variance 
\begin{align}
    \mu\left[f(X_*)\right] &= \Sigma_*^T \left(\Sigma + \sigma^2 \mathbb{1}  \right)^{-1}\mathbf{y}, \\
    \sigma^2\left[f(X_*)\right] &= \Sigma_{**}- \Sigma^T_* \left(\Sigma + \sigma^2 \mathbb{1} \right)^{-1} \Sigma_*. 
\end{align}
The above formulas enable us to make predictions based on the mean value of the Gaussian process at a certain point. Moreover, the variance associated with a prediction at an unseen point depends only on the input values, and their distance, as measured by the kernel function, to the training points. If good convergence is obtained after the optimization procedure, the variance is small around the input points seen in training and larger if the input point is far away from the training points. The length scale at which this happens is determined by the optimized kernel function parameters $l_k$ from Eq. \ref{eq:kernel}.

GPR models are, in theory, universal function approximators \cite{universal_kernels}; however, this result does not prove that our optimization procedure, based on minimizing the marginal likelihood in Eq. \ref{eq:log_marg_likelihood}, is guaranteed to converge as we increase the number of training samples. We therefore typically run each optimization approximately 20 to 100 times with different initial parameter guesses, to get as close as possible to the optimal solution.

Regarding the computational cost of the GPR itself, both the evaluation of the log marginal likelihood in Eq. \ref{eq:log_marg_likelihood} and the prediction formulas require the inversion of the $n \times n$ covariance matrix, where $n$ is the number of training samples, resulting in a complexity of $\mathcal{O}(n^3)$ per optimization step. For the training set sizes used in this work ($n \lesssim 10^3$, see Appendix~\ref{app:exp_details}), this cost is negligible compared to generating the simulated training data itself, even when accounting for the repeated optimizer restarts.

\section{Superconducting Qubit Noise Models}\label{app:noise_models}

In this section we present all the error sources currently limiting the gate fidelities of transmon quantum computers. Here we will also make the distinction between an error \textit{source} and an error \textit{type}. As an example, consider leakage to the higher levels of a transmon during a single-qubit gate operation - we consider this to be an error \textit{type}, which can have several error \textit{sources}, such as non-unitary heating transitions or unitary leakage dynamics due to the small anharmonicity.

The proposed noise model represents a highly detailed simulation of the error process from first principles in transmon devices. Our model improves upon simple pulse-level simulations with static Lindblad jump operators by including the following effects:
\begin{itemize}
\item We improve the modelling of environmental effects in the two-qubit gate scheme. Namely the $T_1$ decay by considering the many-body effects in the non-unitary dynamics derived from the Hamiltonian of a physically realistic environment of two-level systems (TLSs), as well as considering the non-Markovian flux noise as a more realistic description of the pure dephasing of flux-tunable transmons. 
A similarly detailed Hamiltonian modelling of noise in ion trap systems, with good experimental agreement can be found in Ref. \cite{Lotshaw_2022}.
\item Additionally, we have compiled realistic estimates of the accuracy of the control electronics and calibration procedure, focusing on the most common microwave pulse driven single-qubit gates with the DRAG pulse scheme \cite{Motzoi_2009} and the non-adiabatic implementation of the CZ gate in a tunable coupler architecture \cite{Sung_2021,Xu_2021,Yan_2018,google_qec_2022, google_qec_2024}.
\item We include non-Markovian errors such as $1/f$-noise and flux tails, also in the two-qubit gate case.
\end{itemize}

For two-qubit gates we have chosen an implementation that has shown very high fidelities, does not suffer from residual ZZ interaction and has been implemented as the native entangling operation in some of the most significant experiments in the field \cite{google_majoranas_2022,google_qec_2022,google_time_crystals_2021, google_qec_2024} by companies such as Google, with IBM also having recently announced a tunable coupler based chip \cite{IBM_roadmap,stehlik_2021_ibm_tqg}. Additionally, it has been shown that the tunable coupler based CZ gate can also be extended to co-design processors, such as the one presented in Ref. \cite{algaba_2022, renger_2026}.

\subsection{Single-Qubit Gates}\label{app:sqg_noise_model}

We model the transmon as a driven anharmonic oscillator, with a Hamiltonian of the form 
\begin{equation}\label{eq:sqg_ham}
    \hat{H} = \omega \, \hat{a}^\dagger \hat{a} + \frac{\alpha}{2}\hat{a}^\dagger\hat{a}^\dagger \hat{a} \hat{a} - i \Omega(t)\left( \hat{a} - \hat{a}^\dagger\right), 
\end{equation}
where $\hat{a}$ are bosonic annihilation operators, $\omega$ is the frequency of the qubit and $\alpha$ is the anharmonicity. The last term above represents the action of the control waveform parametrized by the function $\Omega(t)$, which is additionally parameterized by the two quadrature controls $I(t)$ and $Q(t)$ so that
\begin{equation}\label{eq:IQ}
\Omega(t) = I(t) \sin\left(\omega_d t \right) + Q(t) \cos\left(\omega_d t \right).
\end{equation}
We accentuate here that the frequency of the drive $\omega_d$ and the transmon transition frequency $\omega$ are not necessarily equal \cite{Krantz_2019}.

The above parametrization is useful when employing the DRAG pulse shape \cite{Motzoi_2009}, which is commonly used due to its effectiveness in reducing leakage to the $f$-state of the transmon, as well as its simplicity and ease of calibration. The DRAG scheme consists of applying any finite pulse shape $s_0(t)$ on one quadrature, and its scaled derivative $s_1(t) \propto \dot{s}_0(t)$ on the other \cite{Motzoi_2009}. More specifically, we use Gaussian pulses defined by
\begin{align}\label{eq:gaussianDRAG}
    s_{0}(t) &= 
        \begin{cases}
        A \exp\left(- \frac{(t - \mu)^2}{2\sigma^2}\right) - B & \text{if } 0 < t < T,\\
        0 & \text{else. } 
    \end{cases}\\
    s_{1}(t) &= - \beta \frac{t - \mu}{\sigma^2} s_0(t).
\end{align}
In other words, $s_0(t)$ is a Gaussian envelope with a finite time cut-off $T$, and $s_1(t)$ is the re-scaled derivative of $s_0(t)$. By using $I(t) = s_0(t)$ and $Q(t) = s_1(t)$, or $I(t) = s_1(t)$ and $Q(t) = s_0(t)$, we can choose between implementing qubit rotations around the $X$ or $Y$-axes of the Bloch sphere.

The parameters $\sigma$, $\mu$ and $T$ are considered to be fixed, while $A$ and $\beta$ are tuned in the calibration procedure. The offset $B$ is set so that the truncated Gaussian curve does not have any discontinuities, i.e. $s_0(t = 0) = s(t = T) = 0$. This definition assumes the pulse is performed at time $t = 0$, however if this is not the case, as we are also interested in performing simulations of multiple gates, the formulas are translated so that $s_{0,\tau(t)} = s_0(t - \tau)$ so that the pulse begins at time $\tau$.

All of our single-qubit gate simulations also include some electronics imperfections such as the finite sampling rate, which we fix to $1/\delta t  = 2.4$ GHz. This means the real output of the electronics is given by discretized values 
\begin{equation}\label{eq:pulse_discretization}
    \tilde{I}(t) = I\left(\left\lfloor \frac{t}{\delta t}\right\rfloor \delta t \right),
\end{equation}
and equivalently for the second quadrature $Q(t)$. While we always assume our pulse is continuous when programming the pulse shapes, the output is actually discretized.

\subsubsection{Amplitude damping errors}\label{sec:sqg_T1}
One of the best-known error sources is amplitude damping characterized by the so-called $T_1$ decay time, which can be easily measured by preparing the qubit in the excited state and measuring the population at different times.

It is believed that for current hardware the observed $T_1$ times are limited by the coupling of the transmon to a bath of two-level systems (TLSs) \cite{Premkumar2021,carroll_2022,Muller_2019,siddiqi_2021_review}. The interaction originates from the coupling of the TLS electric dipole to the transmons charge degree of freedom \cite{Lisenfeld_2019}. There are also other physical contributions to $T_1$, such as interactions with non-equilibrium Bogoliubov quasi-particles; however, their effect was found to be smaller in some studies on transmon qubits \cite{Riste_2013,Kurter2022}, in the absence of burst events \cite{McEwen_2021}.

Except for rare occasions of resonant qubit-TLS interactions, the observed $T_1$ decays are well-described by an exponential curve \cite{Burnett_2019}, and therefore modelled with the Markovian approximation. Motivated by the results of Ref. \cite{Serniak_2015}, where the qubit effective temperature was shown to be considerably higher than the cryostat temperature and can result in excited state populations in the order of 1\% \cite{Heinsoo_2018,Yin_2015,Wenner_2013}, we also consider that the bath producing the $T_1$ decay is at a finite temperature, which means that there is a finite heating rate in the model.

We use the Lindblad equation to simulate the dynamics of such finite temperature $T_1$ decay
\begin{equation}\label{eq:T1_lndb}
    \frac{\mathrm{d}\hat{\rho}}{\mathrm{d}t} = -i \left[\hat{H}, \hat{\rho} \right] + \sum_{i = +,-} \Gamma^i_1 \left( \hat{C}_i \hat{\rho} \hat{C}_i^\dagger + \frac{1}{2}\left\{\hat{C}_i^\dagger\hat{C}_i,\hat{\rho} \right\}  \right),  
\end{equation}
with $\hat{C}_- = \hat{a}$ and $\hat{C}_+ = \hat{a}^\dagger$ and the $T_1$ decay rate in a standard excited state population decay experiment is given by $\Gamma_1 =  \Gamma^+_1 +  \Gamma^-_1$. The ratio of both rates is determined by the detailed balance condition $\Gamma_1^+/\Gamma_1^- = e^{-\omega/k_B T_\mathrm{eff}}$, which is valid if we assume the qubit is embedded in a thermalized environment \cite{breuer_petruccione_oqs}. 
The exponential suppression of the heating rate $\Gamma_1^+$ implies that it has a small effect on the dynamics itself. Nevertheless, it has a significant  effect on the state preparation, as it means that the qubit has some residual excited state population initially. 

In this model, the decay rate from the second to first excited state is exactly twice the rate of the excited to ground state transition, which is in reasonable agreement with the transmon qutrit decay rates reported in Refs. \cite{Goss_2022,Martinez_2022}.

Finally, the above equation is completely independent of the origin of the decay process, as all the characteristics of the environment are reduced to two rates.

\subsubsection{Markovian Pure Dephasing}\label{sec:sqg_Mark_pd}

In the majority of cases, it is observed that the $T_2$ decay time measured in a Ramsey experiment, of a qubit is not limited by $T_1$, or more specifically $T_2 < 2T_1$ \cite{Krinner_2022,stehlik_2021_ibm_tqg}.

Much like the amplitude damping contribution characterized by the decay time $T_1$, we also consider a Markovian contribution to the pure dephasing dynamics of the system. To this aim, we add an additional jump operator to the Lindblad equation in Eq. \ref{eq:T1_lndb}, of the form $\hat{C}_\phi = \hat{a}^\dagger \hat{a}$ with the associated decay rate $\Gamma_\phi$. The observed $\Gamma_2 = 1/T_2$ decoherence rate under this Markovian model is given by $\Gamma_2 = \Gamma_1/2 + \Gamma_\phi $. 

The pure dephasing dynamics can be probed by measuring the pure dephasing decay time of an idling qubit with a Ramsey experiment or by additionally applying a dynamical decoupling sequence such as a spin echo. However, the above Lindblad equation can not describe the effects of the low frequency noise, that is the largest contributor to the pure dephasing and will be introduced in the next section \cite{cywinski_2008}. 

If we assumed that the Lindblad equation accurately describes the dynamics of our system, the $T_2$ time would remain unchanged even if dynamical decoupling sequences are used in the evolution. However, in practice this is often not the case, meaning that a different mechanism is responsible for most of the pure dephasing observed in typical flux-tunable transmons \cite{rower_2023,Bylander_2011,Koch_2007,Ithier_2005}.

\subsubsection{$1/f$-type flux noise}\label{sec:sqg_1/f}

Comparing the results of Ramsey and spin echo experiments in flux tunable transmons, we often observe a discrepancy between these two decay times. This discrepancy immediately implies that the pure dephasing in our system has a sizeable contribution from low frequency or \textit{non-Markovian} noise. This low frequency noise in flux-tunable transmons originates from the coupling to magnetic dipole moments of TLSs in the vicinity of the SQUID loop or possibly also from the flux line \cite{Braumuller_2020}, and typically exhibits a $1/f$ noise power spectrum \cite{siddiqi_2021_review,rower_2023,Ithier_2005}.

For a flux-tunable transmon, with two asymmetric Josephson junctions, the relationship between the qubit frequency and external flux threading the SQUID loop is given by \cite{Koch_2007}
\begin{equation}\label{eq:frq_to_flux}
    \omega(\Phi) = \omega_\mathrm{max} \sqrt{\left| \cos\left(\frac{\pi \Phi}{\Phi_0}\right) \right| \sqrt{1 + d^2 \tan^2\left(\frac{\pi \Phi}{\Phi_0}\right)}},
\end{equation}
where $\Phi_0$ is the flux quantum, and $d$ is the junction asymmetry defined by the two Josephson energies as $d = |E_{J_1} - E_{J_2}|/ (E_{J_1} + E_{J_2})$. The maximal frequency $\omega_\mathrm{max}$ can be expressed in terms of the transmon charging $E_C$ and total Josephson energies $E_{J\Sigma} = E_{J_1} + E_{J_2}$ as $\omega_\mathrm{max} = \sqrt{8 E_C E_{J\Sigma}}$. Flux noise can then be treated as a perturbation of the external flux threading the SQUID loop $\Phi \rightarrow \Phi + \delta\Phi$.

In order to capture these time-correlated non-Markovian effects of the flux noise, we model this noise via a classical stochastic process, comprised of a large number of random telegraph signals. We have chosen this approach over using a completely Gaussian Ornstein-Uhlenbeck process since a random telegraph signal can be interpreted as a classical version of the switching of a single magnetic TLS \cite{Bergli_2009,Muller_2019}.

As mentioned, the noise couples to the qubit via the flux, meaning that the interaction term of the Hamiltonian has the following form \cite{Koch_2007}
\begin{equation}\label{eq:sqg_1/f}
    \hat{H}_{1/f}(t) = \frac{\partial\hat{H} }{\partial \Phi} \, \delta\Phi(t) = \frac{\partial\omega }{\partial \Phi} \delta\Phi(t) \hat{a}^\dagger\hat{a},
\end{equation}
where the above Hamiltonian describes one realization of the stochastic process $\delta\Phi(t)$, which is generated with a $1/f$ noise power spectral density. The flux dispersion of the transmon $\frac{\partial\omega }{\partial \Phi}$ can be simply computed from Eq. \ref{eq:frq_to_flux}.

Since every quantum experiment requires a large number of shots, for each shot the stochastic process $\delta\Phi(t)$ will have a different realization. Therefore, when we average over a number of simulations, each with a different realization of $\delta\Phi(t)$ (with the same statistics), the unitary dynamics of each individual realization results in a pure dephasing decay.

\subsubsection{Calibration errors}\label{sec:sqg_calib}

We have already mentioned that the pulse parameters in Eq. \ref{eq:gaussianDRAG} must be calibrated accordingly to ensure the best possible gate performance. Each of these parameters $A$, $\beta$ as well as the drive frequency $\omega_d$ are tunable and must be optimized for each qubit separately. 

\paragraph{Amplitude} The amplitude of the pulse $A$ must be optimized so that the angle of rotation is set to the correct value. Small deviations will manifest themselves as over or under rotated gates. In the calibration procedure, it is often assumed that the amplitude of a $\pi/2$ rotation is simply half the amplitude of a $\pi$ rotation, i.e. that the relation between the angle of rotation and pulse amplitude is linear \cite{chen_phd_thesis}, which might not be accurate due to non-linearity in the control electronics \cite{Moreira_2022,Lazar_2022}. This assumption results in a mismatched amplitude $A$, as often we wish to calibrate both $\pi/2$ and $\pi$ rotations. We characterize the magnitude of this error by introducing the parameter
\begin{equation}\label{eq:def_epsA}
    \epsilon_A = \varphi \times (A - A_\varphi) / A_\varphi,
\end{equation}
which characterizes the angle of over or under rotation, based on the ideal value for the amplitude $A_\varphi$. This effect was observed to manifest itself in an under rotation angle of up to 3 degrees in Ref. \cite{Lazar_2022}.

An additional source of the discrepancy has also been reported in Ref. \cite{Li_2023}, where the amplitude of the pulse generated by the electronics was observed to fluctuate on the timescale of hours, by up to 0.3 \%. This mismatch would correspond to an angle of approximately $0.5^\mathrm{o}$ or 1$^\mathrm{o}$ for a $\pi/2$ or $\pi$ rotation respectively.

\paragraph{DRAG parameter} The parameter $\beta$ determines the amplitude of the derivative quadrature, and a mismatch will result in increased leakage to the second excited state of the transmon. In the truncated qubit subspace, such an error is non-trace-preserving. We characterize the error in this parameter by considering the relative offset of the value
\begin{equation}\label{eq:def_epsB}
    \epsilon_\beta = (\beta - \beta_\varphi)/ \beta_\varphi, 
\end{equation}
compared to the ideal value $\beta_\varphi $.

\paragraph{Frequency detuning} Perhaps the most complex effect of miscalibrated parameters results from the frequency mismatch of the drive, i.e. when $\omega_d - \omega \neq 0$. This mismatch results in a phase difference between the qubit and the drive which grows with time. This can be easily understood if we simplify our system with the following assumptions. We consider only the first two-levels of the transmon driven with a simplified pulse on only one quadrature, i.e. $I(t) = s_0(t)$ and $Q(t) = 0$ with the pulse being applied at time $\tau$, and perform the rotating wave approximation to obtain the driving Hamiltonian in the rotating frame of the qubit, in the basis $\{|0\rangle , |1\rangle \}$ 
\begin{equation}
    \hat{H}_{\delta \omega}(t) = -\frac{s_0(t - \tau)}{2} 
    \begin{pmatrix}
    0 & e^{it (\omega - \omega_d) }\\
    e^{- it (\omega - \omega_d) } & 0
    \end{pmatrix}.
\end{equation}
This driving Hamiltonian can be interpreted as a rotation around an axis in the $x$-$y$ plane of the Bloch sphere which slowly drifts in time with an angular frequency $\delta \omega = \omega - \omega_d$. While detunings large enough to have an effect on a timescale of  single gate time might not be realistic, if we are performing a longer algorithm the single qubit rotation at time $\tau$ will be shifted by an angle of $\delta \omega\, \tau$. While this error does not have any history dependence, its magnitude therefore depends on the time of the execution of the gate. In the above formula we have also implicitly defined the qubit Bloch sphere such that the drive and qubit are in phase at time $t = 0$. Since this may not always be the case, we introduce an additional error parameter $\epsilon_\mathrm{phase}$ to characterize the initial phase difference between the drive and the qubit.

Realistically, since measuring the qubit frequency via a Ramsey experiment is very accurate, such effects might be a consequence of the qubit frequency drifts due to the stochastic nature of the TLS environment of the transmon \cite{bejanin_2021}. These drifts were observed to be in the range of a couple of kHz with infrequent jumps up to 20 kHz \cite{Burnett_2019}.

\subsection{Two-Qubit Gates}\label{app:tqg_noise_model}

We consider a non-adiabatic CZ gate based on the two-qubit gate scheme using tunable couplers that was analyzed in Refs. \cite{Sung_2021,Xu_2021,Yan_2018,chu_2021} with similar schemes proposed in Refs. \cite{Marxer_2022,Sete_2021,Goto_2022,stehlik_2021_ibm_tqg}.

This gate is implemented by introducing another non-computational element into the circuit known as a coupler ($C$). The two computational transmons ($Q_{1,2}$) are then capacitively coupled to the coupler and to each other. The coupler is also a flux-tunable transmon; however, in the idling configuration the frequency of this element must be significantly detuned from the frequency of both computational transmons to suppress the interaction between them.

Such a system is modelled with the following Hamiltonian,
\begin{align}\label{eq:tqg_ham}
    \hat{H} &= \sum_{i \in \{Q_1,C,Q_2\}} \omega_i \hat{a}_i^\dagger \hat{a}_i + \frac{\alpha_i}{2} \hat{a}_i^\dagger\hat{a}_i^\dagger \hat{a}_i \hat{a}_i \nonumber\\
    &- \sum_{\substack{i,j \in \{Q_1,C,Q_2\} \\ i\neq j}} g_{ij} (\hat{a}_i^\dagger - \hat{a}_i) (\hat{a}_j^\dagger - \hat{a}_j)
\end{align}

The CZ gate is implemented by tuning the frequency of the coupler via a flux pulse closer to the frequency of the computational transmons, so that the coupler frequency is a time dependent function $\omega_C(t)$. The couplings between the transmons $g_{ij}  = \beta_{ij} \sqrt{\omega_i \omega_j}$ actually also depend on the frequencies, meaning that while $g_{Q_1 Q_2}$ is constant, $g_{Q_{1,2} C}$ is also time dependent. The prefactors $\beta_{ij}$ depend on the coupling capacitances, as well as self-capacitances of the transmons in the lumped element circuit model.

We use a flattop-Gaussian pulse in the realization of our gates. This pulse shape is the result of a convolution of a rectangular pulse and a Gaussian function, with the equation
\begin{equation}\label{eq:flattop_gaussian}
    f(t) = \frac{1}{2}\left[\mathrm{erf}\left(\frac{t - \tau_b}{\sqrt{2}\sigma}\right) - \mathrm{erf}\left(\frac{t - \tau_b  - \tau_c}{\sqrt{2}\sigma}\right) \right],
\end{equation}
where $\tau_c$ is the duration of the rectangular pulse, $\sigma$ is the standard deviation of the Gaussian, $\tau_b$ is the rise time of the pulse and erf$(\cdot)$ is the error function. The above formula still represents a pulse with infinite duration and fixed amplitude, so the actual coupler frequency has a time dependence of $\omega_C^\mathrm{idle} + \omega_C(t)$, with
\begin{equation}\label{eq:coupler_pulse}
    \omega_C(t) = 
    \begin{cases}
        A\left(f(t) - B\right) & \text{if } 0 < t < \tau_c + 2\tau_b,\\
        0 & \text{else. } 
    \end{cases}
\end{equation}
The constant $A$ represents the amplitude of the pulse and the offset $B$ is there to avoid any discontinuities in the coupler frequency due to the finite duration of the pulse. In Eq. \ref{eq:coupler_pulse}, we have set the gate duration to $\tau_c + 2\tau_b$, which makes the pulse symmetric and fixes $B = f(0)$. The pulse in the above parametrization starts at time $t = 0$ and must be accordingly shifted if we choose to perform a pulse at a different time.

Since the convolution of two functions is simply the product of their Fourier transforms, the Gaussian function suppresses the slowly decaying frequency tail of the rectangular function, thus minimizing the probability of exciting higher states.

We have also considered the effects of finite sampling of the electronics with a rate of 2.4 GHz as in Eq. \ref{eq:pulse_discretization}, however after the flux pulse is passed through a low-pass filter with a cut-off of 1 GHz, as in Ref. \cite{Marxer_2022}, almost no difference compared to the analytical pulse shape in Eq. \ref{eq:flattop_gaussian} was observed.

As the coupler flux pulse is being applied, the level repulsion involving the levels of the double excitation manifold results in an avoided crossing between the states $|1_{Q_1} 0_C 1_{Q_2}\rangle \leftrightarrow |0_{Q_1} 0_C 2_{Q_2}\rangle$, where the subscripts refer to the qubits ($Q$) or the coupler ($C$). In this formulation, the frequency of $Q_2$ is larger than the frequency of $Q_1$. The population of the computational $|1_{Q_1} 0_C 1_{Q_2}\rangle $ state then undergoes a full Rabi oscillation which implements a conditional phase \cite{chu_2021}. 

The benefits of implementing the gate in this specific manner are two-fold. First of all, unlike having directly coupled computational transmons where the interaction strength asymptotically decays to zero with the qubit-qubit detuning, a well-designed system with a tunable coupler has one or two coupler frequencies where the interaction between the qubits can be tuned to \textit{exactly} zero in theory \cite{Yan_2018}, with an effective coupling strength of approximately 1 kHz reported in Ref. \cite{Sung_2021}. Furthermore, since the gate is based on a non-adiabatic interaction the gates are relatively fast, typically on the timescale of 30 to 100 ns \cite{Yan_2018,Sung_2021,Marxer_2022,chu_2021,stehlik_2021_ibm_tqg}. 

The introduction of a non-computational element such as the coupler is a large issue for the characterization of the system. The coupler, which has its own error sources, cannot be read out, and neither are we able to apply microwave pulses, meaning that we can only perform $z$-axis rotations via flux tuning. Adding additional control or readout lines to the chip should be avoided due to the increased risk of cross-talk, as well as scalability issues arising from the increased heat load of additional control lines in the cryostat. Moreover, the population of the computational $|11\rangle$ state is transferred outside of the computational subspace during the gate duration, and while measurements of the second excited state population are doable, they are typically not implemented as they are not needed for computation. All in all, this means we must characterize a system without being able to measure or control a significant part of the Hilbert space.

Additionally, the computational basis is now given by the eigenstates of the full Hamiltonian in Eq. \ref{eq:tqg_ham}, and is therefore slightly delocalized due to the coupling between the transmons \cite{chu_2021}. We identify the computational state eigenbasis via maximum overlap with the local basis, so that the computational $|ij\rangle$ state, where $i,j \in \{0,1\} $, is the eigenstate $|\psi\rangle $ of Eq. \ref{eq:tqg_ham} which maximizes the overlap $\left| \langle \psi | i_{Q_1} 0_C j_{Q_2}\rangle \right|^2$. The basis states are therefore defined with the coupler in the ground state. This definition of the computational basis is valid only when the coupler is significantly detuned from the qubit frequencies, and the basis can be uniquely defined, as the overlaps in this regime are close to one.

\subsubsection{Amplitude damping errors}\label{sec:tqg_amplitude_damping}

Much like the single-qubit case, errors related to Markovian $T_1$ decay are still a major source of infidelity for the two-qubit gate system. 

The simplest approach to model this type of incoherent dynamics would be to consider a Lindblad equation with the same jump operators as in \ref{eq:T1_lndb}, but localized to each transmon. Neglecting any heating transitions, it is obvious that the steady state of such a system is simply the state $|0_{Q_1}0_C 0_{Q_2}\rangle$. It can easily be checked that this is not the ground state of the Hamiltonian with non-zero coupling coefficients $g_{ij}$, and therefore such a model is flawed when trying to describe a system in thermodynamic equilibrium with its environment. 

While this local approximation to the Lindblad dynamics tends to be accurate in the dispersively coupled regime, the diabatic two-qubit gate is operated in the strongly coupled or near-resonant regime, where the eigenstates of the system are significantly delocalized from the physical transmon states \cite{chu_2021}. In this regime, the local approximations will not accurately describe the dynamics.

We obtain a more accurate description by following a microscopic derivation of the Lindblad equation \cite{breuer_petruccione_oqs} under the assumptions that each transmon is coupled to its own bath. This is supported by the idea that the main source of $T_1$ decay are TLSs with electric dipoles, and the electric field of each transmon is strongest in its immediate vicinity. Significant electromagnetic interactions between transmons (beyond the implemented capacitive couplings) would result in large crosstalk and therefore dysfunctional qubits. We therefore derive a Lindblad equation where each transmon is coupled to a number of TLSs, similarly to the standard tunnelling model analyzed in Ref. \cite{cho2022,Muller_2019}.

The resulting \textit{global} Lindblad equation that describes the non-unitary decay of a coupled multi-body system preserves the Lindblad form, but with jump operators that couple the eigenstates of the system. This can be written down as

\begin{align}\label{eq:tqgT1_lndb}
    \frac{\mathrm{d}\hat{\rho}}{\mathrm{d}t} &= -i \left[\hat{H} + \hat{H}_{LS}, \hat{\rho} \right] \nonumber \\
    &+ \sum_{a,b} \Gamma^{ab}_1 \left( \hat{C}_{ab} \hat{\rho} \hat{C}_{ab}^\dagger + \frac{1}{2}\left\{\hat{C}_{ab}^\dagger\hat{C}_{ab},\hat{\rho} \right\}  \right),  
\end{align}
where the density matrix $\hat{\rho}$ refers to the whole system of three transmons. The jump operators represent transitions between eigenstates (denoted with indices $a$ and $b$), and are defined as $\hat{C}_{ab} = | a\rangle \langle b|$. The Lamb-shift Hamiltonian is denoted with $\hat{H}_{LS}$.

A global model is needed to explain the $T_1$ times observed during the operation of a non-adiabatic CZ gate in Refs. \cite{marxer_2026, Yuan_2020, Marxer_2022}, where the hybridization of the computational states with the coupler results in a reduced decay time during the operation of the gate, thus further discouraging the use of local jump operators together with the $T_1$ times measured at the idling point of the system when modelling the incoherent dynamics.

The details of this derivation describing the calculation of the rates $\Gamma_{ab}$, based on the coupling to a TLS bath, can be found in Appendix \ref{app:t1_model}. Identically as in the single qubit case we also include heating transitions via the detailed balance condition, similarly to the model used in Ref. \cite{google_qec_2022, google_qec_2024}. We assume, as an approximation, that $T_1$ is independent of qubit frequency, since the most commonly observed frequency dependence actually exhibits seemingly random fluctuations due to resonant couplings to specific TLSs in the environment \cite{cho2022,carroll_2022}.

It is also worth noting that, while the Lindblad equation considered here describes a Markovian evolution of the full system, the global approach will induce incoherent leakage transitions outside of the computational subspace. Thus, the evolution of the computational subspace in this case can actually be non-Markovian. This can be seen if we consider the transition from the qubit to coupler excited state $|0_{Q_1}0_C1_{Q_2}\rangle \rightarrow |0_{Q_1}1_C0_{Q_2}\rangle$ as an example. If the coupler excited state does not instantaneously decay into the ground state of the system, the presence of coupler excitations will distort the energy levels of the computational basis and therefore negatively affect subsequent gate operations. Since the population of the coupler excited state depends on the coherence times as well as the number of gates performed in the past, the latter history dependence means that this error process is non-Markovian.

\subsubsection{$1/f$-type flux noise}\label{sec:tqg_1/f}

In a similar way as in the single-qubit case, the slowly varying magnetic flux noise due to spin defects or classical electronics is also present in two-qubit gates.

This noise is even exacerbated in the tunable coupler, where the pure dephasing times are observed to be up to an order of magnitude lower compared to the computational transmons \cite{Marxer_2022}. This is a natural consequence of the different design of the coupler, which must be easily tunable over a large frequency range, given the typical trade-off between noise and control.

By making the adiabatic approximation, i.e. assuming that the noise varies slowly enough not to induce any transitions between the eigenstates, which is a justified assumption due to the large magnitude of the low-frequency part of the spectrum, the susceptibility of the coupled system to the slowly varying flux can be characterized by generalizing the single-qubit formula from Eq. \ref{eq:sqg_1/f} so that
\begin{equation}\label{eq:tqg_1/f}
    \hat{H}_{1/f} = \sum_{i \in \{Q_1, C, Q_2 \}} \sum_a  \frac{\partial\varepsilon_a}{\partial\omega_i }\frac{\partial \omega_i}{\partial\Phi_i } \delta\Phi_i(t) | a \rangle \langle a |,
\end{equation}
where we have written the multi-body Hamiltonian from Eq. \ref{eq:tqg_ham} in its eigenbasis with eigenstates $|a\rangle$ and corresponding eigenergies $\varepsilon_a$. Besides the regular flux susceptibility of each transmon frequency $\frac{\partial \omega_i}{\partial\Phi_i }$, the coupling of the system (as well as the dependence of the coupling coefficients $g_{ij}$ on the frequencies) is reflected in the first coefficient of the form $\frac{\partial\varepsilon_a}{\partial\omega_i }$. In the uncoupled case with $g_{ij} = 0$, this coefficient can only be integer or zero, and the above equation reduces to the single qubit case as in Eq. \ref{eq:sqg_1/f}. During the gate operation, in the highly hybridized regime, this coefficient then takes into account the effect of the local flux noise on the hybridized system.

Since the coupler frequency is being tuned during the operation of the gate, this will also affect the flux dispersion of the coupler $\partial \omega_C / \partial \Phi_C$, as seen from Eq. \ref{eq:frq_to_flux}.

The noise is again assumed to be localized to each transmon without any correlation between them. For the noise from the classical electronics this is obvious since each transmon is coupled to its own flux-line with very little crosstalk between them, and the magnetic TLSs are assumed to produce local noise as per the same arguments as for the charge noise. Since the computational basis consists of hybridized eigenstates \cite{marxer_2026, Yuan_2020, Marxer_2022}, the coupler flux noise in particular will result in correlated pure dephasing of the computational states.

Identically as with a single qubit, many realizations of the classical stochastic processes $\delta\Phi(t)$ must be simulated and averaged over uniformly. 

The same model was used to explain the pure dephasing in a tunable coupler setup already in Ref. \cite{chu_2021} and tested out previously in Ref. \cite{Marxer_2022} with good agreement observed.

\subsubsection{Calibration errors}\label{sec:tqg_pulse}
While the flattop-Gaussian pulse from Eqs. \ref{eq:flattop_gaussian} and \ref{eq:coupler_pulse} has a number of free parameters, only two of them need to be tuned to calibrate the gate.

Typically, the rise time $\tau_b$ and standard deviation $\sigma$ are fixed in such a way that $\tau_b > \sigma$, and the amplitude $A$ and rectangular pulse duration $\tau_c$ are tuned. 

As mentioned previously, the gate is based on a Rabi oscillation between the levels $\ket{1_{Q_1} 0_C 1_{Q_2}}$ and $\ket{0_{Q_1} 0_C 2_{Q_2}}$, so the most obvious consequences of imperfect calibration are either the population not completing the full Rabi cycle, resulting in leakage to the $\ket{0_{Q_1} 0_C 2_{Q_2}}$ state or the population returning with the wrong conditional phase. Additionally, errors can occur in the initial and final phases of ramping the pulse up and down. The errors here are more unpredictable, however typically we observe Landau-Zener transitions to other non-computational states. This error occurs even if the pulse is perfectly calibrated and can only be eliminated by choosing a different, more optimal, pulse shape \cite{chu_2021}.

Furthermore, since the flux pulse shifts the energy levels of the hybridized system during the gate, each computational qubit accumulates a deterministic single-qubit phase, which is corrected after the gate with virtual Z rotations. A miscalibration of these correction angles leaves a residual coherent single-qubit phase error, which we characterize by the offset angle $\Delta \Phi_Z$ per gate for each qubit.

\subsubsection{Pulse distortion}\label{sec:tqg_flux_distortion}

Since no waveform generator is perfect and the pulse passes through a number of filters before reaching the transmons, certain pulse distortions have been observed.

First of all, the frequency of the coupler is tuned via a flux pulse. The relationship between the external flux threading the SQUID loop and the coupler frequency was described in Eq. \ref{eq:frq_to_flux}. Typically, the parameter $d$ of the coupler transmon is designed to be either zero or very small to ensure a larger tuneability; however, fabrication inaccuracies may cause deviations from design values. Assuming the coupler is designed with $d=0$, current fabrication inaccuracies might result in $d \sim 10^{-2}$ or even $d \sim 10^{-3}$ with the use of laser annealing \cite{hertzberg_2021}. Since this translates into $d^2 \sim 10^{-4} - 10^{-6} $ and gate operation is done in the regime where the tangent term does not diverge, we neglect the asymmetry term from now on.

We model any flux distortion errors with the following formula.
\begin{equation}
    \Phi(t) = \int_0^t \mathrm{d}t' \, \tilde{\Phi}(t') \mathcal{K}(t - t'),
\end{equation}
where $\Phi(t)$ is the distorted flux and $\tilde{\Phi}(t')$ is the desired flux pulse train, meaning that the functions span over a time period of the whole algorithm. The desired flux distortion is convolved with a kernel function $\mathcal{K}(t - t')$ in a time-local way so that the current distorted flux depends only on the past and not the future.

Based on the measurements of the step response of the electronics in Refs. \cite{Roi_2019_netzero,Sung_2021} we parametrize the kernel function in the following way
\begin{equation}
    \mathcal{K}(t - t') = \delta(t - t') + \sum_n A_n e^{-\frac{|t - t'|}{\tau_n}},
\end{equation}
where the delta function in the first term corresponds to a perfect pulse, i.e. $\Phi(t) = \tilde{\Phi}(t)$ if $A_n = 0\, \forall n$ and we use a handful of exponential tail distortions with a wide range of typical parameters. Each distortion is parametrized by its amplitude $A_n$ and timescale $\tau_n$. While the amplitude must be relatively small, $A_n \lesssim 0.01$, to achieve reasonably high fidelities, the timescales $\tau_n$ can cover a wide range from 10 ns up to 1 $\mu$s and possibly longer \cite{Sung_2021}.

The distortions with timescales significantly exceeding one gate time will not distort a single pulse shape, but will rather offset the coupler frequency during the idling period away from the frequency where the residual $ZZ$-interaction is zero. The magnitude of this offset will depend on the exact value of the time correlation coefficient $\tau_n$ and the number of pulses performed within a time frame specified by this parameter. This makes this error completely history dependent and therefore non-Markovian. 

This offset manifests itself as an unwanted conditional phase during idling, as well as a miscalibrated pulse. The miscalibration arises because we still perform a pulse of the same amplitude, even though the initial coupler frequency has been offset. If the correlation timescale of the distortion is shorter or comparable to the gate duration, it will result in a deformation of the pulse shape, causing miscalibration errors as well as more unwanted leakage transitions during the ramping up and down of the pulse. However, such an error will also have more Markovian behaviour, since the memory of the flux distortion is shorter.

If $\tau_n$ is comparable to a gate duration, a single pulse will rise and fall more slowly than expected and the non-Markovianity of such an evolution is smaller.

\subsection{State preparation}\label{app:state_preparation}
We have already mentioned in Appendix~\ref{sec:sqg_T1} that transmon residual excited state populations are routinely observed to significantly exceed what is predicted by the Maxwell-Boltzmann distribution with the cryostat temperature, which is typically around 15 mK.

Besides the heating transitions in the Lindblad equations \ref{eq:T1_lndb} and \ref{eq:tqgT1_lndb}, if we wish to simulate a realistic experiment, the residual excited state population should be taken into account in the initial state preparation. Therefore, for realistic experiments, we assume that the initial state is a thermal state
\begin{equation}
    \hat{\rho}(t = 0) = \exp{\left(-\beta \hat{H}\right)}/Z
\end{equation}
with an effective temperature $T_\mathrm{eff}$, specified by $\beta = 1/(k_B T_\mathrm{eff})$, typically in the range of 50 mK \cite{Heinsoo_2018,Yin_2015,Wenner_2013}, and $\hat{H}$ is either a single or two-qubit Hamiltonian from Eqs. \ref{eq:sqg_ham} or \ref{eq:tqg_ham}. The partition function of the system enforcing normalization is denoted with $Z$. This state is also the steady state solution of the Lindblad equations \ref{eq:T1_lndb} and \ref{eq:tqgT1_lndb}.

\subsection{Measurement Errors}\label{app:measurement}

We assume standard dispersive measurements of the transmons in the computational subspace. Even though the same setup could also accommodate the readout of the higher excited states, this readout is typically not calibrated and we assume we have no access to it. Additionally, as mentioned previously, the coupler is not connected to any readout resonator. In the more general two-qubit case, the probability distribution of measuring the state $k$ is given by
\begin{equation}\label{eq:measurement}
    P(k|\hat{\rho}) = \frac{\mathrm{tr} \left\{ \hat{P}_k  \hat{\rho}_q \right\}}{\mathrm{tr} \left\{  \hat{\rho}_q \right\}}.
\end{equation}
Here, the density matrix $\hat{\rho}_q$ is defined in the computational subspace, meaning that it is extracted from the density matrix of the full system $\hat{\rho}$. This procedure is described in Appendix~\ref{app:gate_performance} in more detail. Equivalently, the projector $\hat{P}_k$ is also defined in the computational subspace.

The normalization in Eq. \ref{eq:measurement} is needed due to leakage to the excited states of the transmon, so that the probability distribution $P(k|\hat{\rho})$ is normalized. Once the probability distribution is obtained, we sample $N_\mathrm{shots}$ results from it, to describe the finite sampling effects. After this procedure we obtain a vector of probabilities of measuring each state (or bit string) $\mathbf{P}(\hat{\rho}) = (P(00|\hat{\rho}),P(01|\hat{\rho}),P(10|\hat{\rho}),P(11|\hat{\rho}))^T $.

As readout errors in current hardware are large and unavoidable, we model them by transforming the obtained vector of probabilities with a misclassification matrix of the form
\begin{equation}\label{eq:meas_err}
    \tilde{P}(k|\hat{\rho}) = \sum_{i \in \{ 00,01,10,11\}} P(k|i)P(i|\hat{\rho}),
\end{equation}
together with the preservation of probability $ \sum_k P(k|i) = 1$.

\section{Efficient simulations}\label{app:simulations}

\begin{figure}[t]
\centering
\includegraphics[width=.45\textwidth]{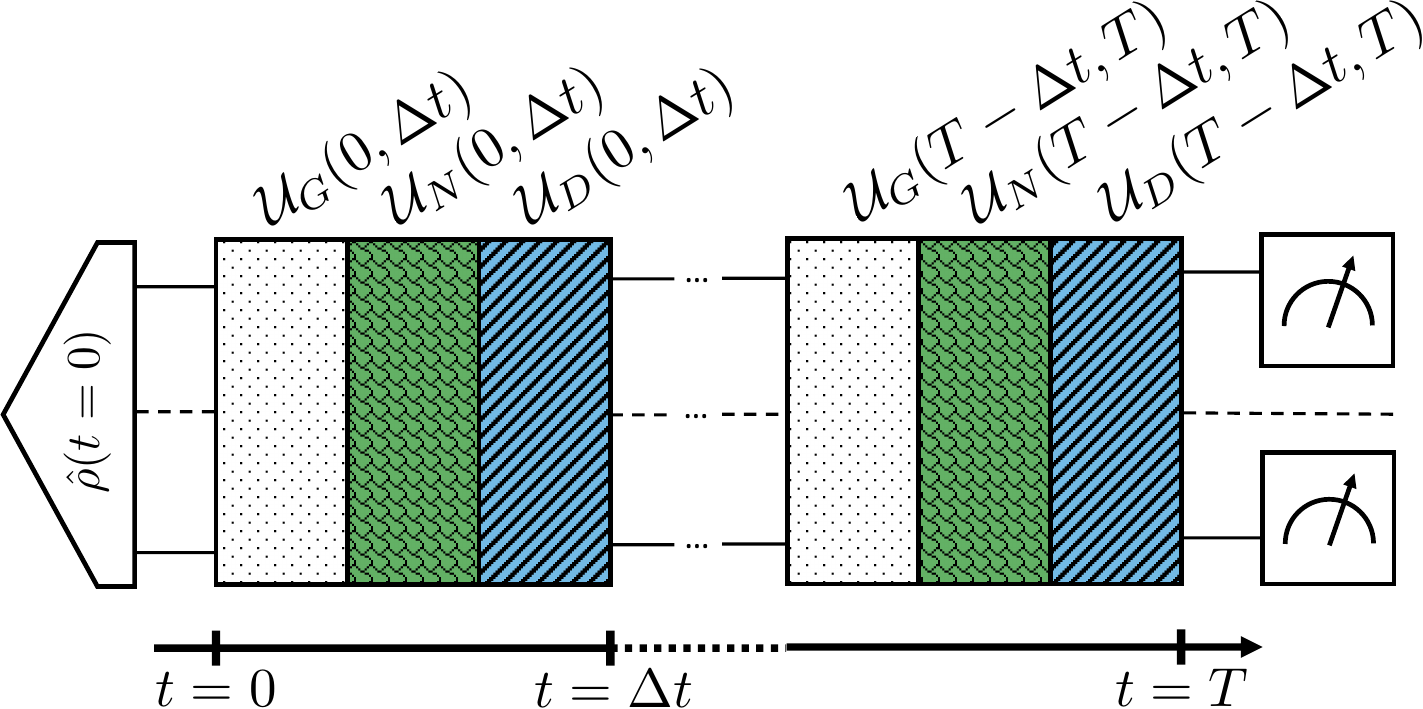}
\caption{\label{fig2:simulation} A schematic representation of Eq. \ref{eq:simulation}, showing how the initial density matrix is subsequently multiplied by the propagators of each noise. What is not included in the figure is the fact that due to the non-Markovian component the result has to be averaged over a large number of realizations of $\mathcal{U}_N$.}
\end{figure}

In the previous section, we have described a variety of error sources, each with its own characteristics, from sources of leakage, coherent, non-unitary and non-Markovian errors. Since we would like to simulate a gate with all these different errors, and possibly add more in the future, or even perform parameter sweeps, one must approach this problem with more careful consideration. 

Since solving noisy pulse level dynamics always reduces to solving some sort of master equation, the addition of $1/f$ noise introduces another degree of complexity due to the needed averaging over many, typically at least hundreds of trajectories. Moreover, solving each trajectory involves solving a stochastic Schr\" odinger equation together with non-unitary dynamics. 

We therefore describe below how we simulate the time dynamics of the quantum system of interest, according to the noise model described in Appendix~\ref{app:noise_models}.

To simplify the evolution, we first vectorize the density matrix of the full system, so that the operator $\hat{\rho}$ is transformed into a vector $\hat{\rho} \rightarrow |\rho\rangle\rangle$, and the more general master equation Eq. \ref{eq:tqgT1_lndb} in Lindblad form is reduced to
\begin{equation}\label{eq:general_vectorrized_evol}
    \frac{\mathrm{d} }{\mathrm{d}t}|\rho(t)\rangle\rangle = \mathcal{L}|\rho(t)\rangle\rangle.
\end{equation}
The form of the superoperator $\mathcal{L}$ then depends on the exact vectorization employed. Perhaps the simplest method is to stack the rows of the density matrix, which results in the Liouvillian superoperator $\mathcal{L}$ with the matrix form
\begin{align}
    &\mathcal{L} = -i\left(\hat{H}\otimes \mathbb{1} - \mathbb{1}\otimes \hat{H} \right) \nonumber \\
    &+ \sum_{a,b} \Gamma_1^{ab} \left(\hat{C}_{ab} \otimes\hat{C}_{ab}^* - \frac{1}{2} \left[\hat{C}_{ab}^\dagger \hat{C}_{ab}  \otimes \mathbb{1} +  \mathbb{1} \otimes \hat{C}_{ab}^\mathrm{T}\hat{C}_{ab}^*  \right]  \right).
\end{align}

Once we add the stochastic contribution due to $1/f$ noise from Eqs. \ref{eq:sqg_1/f} and \ref{eq:tqg_1/f} to the evolution, we are forced to solve a master equation a large number of times with different realizations of the classical stochastic process. However, for all practical purposes the noise in the system is weak, and the evolution can be separated into disjoint parts \cite{abad2022}.

We take this into account by discretizing the evolution into smaller time steps $\Delta t$, and approximating each time step with a first order truncated propagator. More formally, since the general solution to Eq. \ref{eq:general_vectorrized_evol} with a time dependent $\mathcal{L}(t)$ can be written in terms of a generalized propagator $\mathcal{U}(t_2, t_1)$
\begin{equation}
    |\rho(t)\rangle\rangle = \mathcal{U}(t, 0)|\rho(0)\rangle\rangle,
\end{equation}
we can further decompose the superoperator $\mathcal{L}(t) = -i \mathcal{G}(t) - i \mathcal{N}(t) + \mathcal{D}(t)$ into three separate parts:
\begin{itemize}
    \item $\mathcal{G}(t)$ - representing the closed system dynamics of the system corresponding to the superoperator form of the Hamiltonians in Eq. \ref{eq:sqg_ham} or \ref{eq:tqg_ham}.
    \item $\mathcal{N}(t)$ - corresponding to a single realisation of the classical $1/f$ noise process from Eqs. \ref{eq:sqg_1/f} and \ref{eq:tqg_1/f}.
    \item $\mathcal{D}(t)$ - corresponding to the non-unitary dynamics of any jump operators in the Lindblad equations \ref{eq:T1_lndb} and \ref{eq:tqgT1_lndb}.
\end{itemize}
We can proceed to write down the formal solution of Eq. \ref{eq:general_vectorrized_evol} with a Dyson series expansion
\begin{widetext}
\begin{align}
    \mathcal{U}(t, 0) &= \prod_{n=0}^{\lfloor t/\Delta t \rfloor} \mathcal{U}([n+1]\,\Delta t,  n\, \Delta t) \\
    &=  \prod_{n=0}^{\lfloor t/\Delta t \rfloor} \left[ \mathbb{1} + \int_{n\, \Delta t}^{(n+1)\,\Delta t} \mathrm{d}t' \mathcal{L}(t') + \int_{n\, \Delta t}^{(n+1)\,\Delta t}\mathrm{d}t' \int_{n\, \Delta t}^{t'}\mathrm{d}t'' \mathcal{L}(t')\mathcal{L}(t'') + \cdot\cdot\cdot \right] \\
    &= \prod_{n=0}^{\lfloor t/\Delta t \rfloor} \left[ \mathcal{U}_G([n+1]\,\Delta t,  n\, \Delta t) -i \int_{n\, \Delta t}^{(n+1)\,\Delta t} \mathrm{d}t' \mathcal{N}(t') + \int_{n\, \Delta t}^{(n+1)\,\Delta t} \mathrm{d}t' \mathcal{D}(t') + \cdot\cdot\cdot \right] \\
    &\approx \prod_{n=0}^{\lfloor t/\Delta t \rfloor} \mathcal{U}_G([n+1]\,\Delta t,  n\, \Delta t)\mathcal{U}_N([n+1]\,\Delta t,  n\, \Delta t) \mathcal{U}_D([n+1]\,\Delta t,  n\, \Delta t). \label{eq:simulation}
\end{align}
\end{widetext}
In the second to last step we have replaced the linearized propagator with the full propagator of the unitary generator $\mathcal{G}(t)$. This is done since it significantly lowers the error of the expansion as the magnitude of the elements of the matrix form of $\mathcal{G}(t)$ is much larger compared to the noise generators $\mathcal{N}(t)$ and $\mathcal{D}(t)$. The leading error terms of a single step with duration $\Delta t$ are therefore $\propto \int_{n\, \Delta t}^{(n+1)\,\Delta t}\mathrm{d}t' \int_{n\, \Delta t}^{t'}\mathrm{d}t'' \left[ \mathcal{G}(t')\mathcal{D}(t'') + \mathcal{G}(t')\mathcal{N}(t'') + \cdot\cdot\cdot\right]$, resulting in a global evolution error of $\mathcal{O}( \Gamma \omega T \Delta t)$, where $\Gamma$ is either the pure dephasing or amplitude decay rate, $\omega$ is the largest frequency of the system, $T$ is the duration of the simulation and $\Delta t$ is the duration of a single time step. While the magnitude of the effect of the noise is of the order of $\varepsilon \sim \Gamma_1 T$ for amplitude decay and $\varepsilon \sim (\Gamma_\phi T)^2$ for the $1/f$-noise, this means that for accurate results $\omega \Delta t \ll \varepsilon$. 

The formula in Eq. \ref{eq:simulation} basically splits the evolution first into a large number of small timesteps and each time step into a separate contribution of each environmental noise source. In the last line we have also used the full propagator for this noise, instead of the approximate linearized form. This is done to avoid unphysical properties of the density matrix.

Splitting the noisy and unitary evolutions is a very common approximation when simulating \cite{Georgopoulos_2021} and benchmarking \cite{blume-kohout_2022_taxonomy,Wallman_2018} current noisy devices, even though the error scaling is not very favourable. Here we have gone one step further, by also including the effect of the noise during the gate operation.

The main computational advantage of this is that the set of propagators of the unitary and Lindblad dynamics $\mathcal{U}_G$ and $\mathcal{U}_D$ are always identical, while the generators $\mathcal{U}_N$ depend on each trajectory of the $1/f$-noise we are considering. This means that instead of solving the master equation for each trajectory, we just need to generate the propagators $\mathcal{U}_G$ and $\mathcal{U}_D$ by propagating $D^2$ linearly independent states, where $D$ is the dimension of the system. Since the $1/f$ noise acts on the eigenbasis of the system, we need to diagonalize the Hamiltonian once per each time step and afterwards the exponentiation of the matrix is trivial. Typically we want to average over approximately $N_\mathrm{traj} \sim 1000$ trajectories, with a system of dimension $D$. Using $N_{\Delta t} = \lfloor T/\Delta t \rfloor$ time steps, the complexity of the evolution is: 
\begin{enumerate}
    \item Propagate $D$ linearly independent \textit{pure} states to obtain the generators $\mathcal{U}_G$, corresponding to the white dotted area in Fig. \ref{fig2:simulation}
    \item Propagate $D^2$ linearly independent states to obtain the generators $\mathcal{U}_D$, corresponding to the blue diagonal patterned area in Fig. \ref{fig2:simulation}.
    \item Multiply the propagators $\mathcal{U}_G$ and $\mathcal{U}_D$, corresponding to $N_{\Delta t}$ multiplications of $D^2\times D^2$ matrices. 
    \item Diagonalize the Hamiltonian of the system at each time step, corresponding to $N_{\Delta t}$ diagonalizations of a $D\times D$ matrix.
    \item Generate the propagators $\mathcal{U}_N$, corresponding to the green area in Fig. \ref{fig2:simulation}, by simple numerical integration in the eigenbasis and then transform the generator into the correct frame, resulting in 2 matrix multiplications of size $D\times D$ per time step, all together $2N_{\Delta t}$ matrix multiplications per trajectory.
    \item Once all the propagators are generated, we need to multiply the initial state at each time step with the precomputed double propagator $\mathcal{U}_{G,D}$ and then with the corresponding propagator $\mathcal{U}_N$, meaning that we need to multiply a $D^2\times D^2 $ matrix with a $D^2$ dimensional vector twice for each trajectory. 
\end{enumerate}
The most numerically difficult steps depending on the exact size of the system are the matrix multiplications in step 3. with a complexity of $\mathcal{O}(N_{\Delta t}D^6)$ and the actual propagation in step 6. with complexity $\mathcal{O}(N_\mathrm{traj} N_{\Delta t}D^4)$.

\section{Gate Performance Metrics}\label{app:gate_performance}

The first step in benchmarking a quantum gate is to define a performance measure, which is typically a distance measure between the desired and observed outputs of the gate. The most common measures include state or process fidelities, trace distance, or diamond norms \cite{nielsen_chuang}.

When considering contributions of individual error sources we limit ourselves to the effects of a single error source being present. This means that we do not independently simulate the performance of the gate for arbitrary combinations of errors, which would also take into account any potential interplay between them. 

We have shown in the previous section that many of the error sources affecting quantum gates may be non-Markovian or time dependent, meaning that the contribution of such an error source is time or history dependent and cannot be condensed into a single number. A gate performance measure that would take into account such effects has therefore not yet been defined, and is beyond the scope of this work.

Instead, we choose to monitor the evolution of the averaged state fidelity after applying a series of gates. For non-unitary Markovian noise sources such a contribution will increase linearly and for coherent errors quadratically provided the error is small. More complex errors might display more complex behaviour, with non-monotonic decays being associated with non-Markovianity \cite{Figueroa-Romero_2021}. 

Since the simulations are based on state propagation, we choose to examine the gate performance in terms of the averaged state fidelity defined as
\begin{equation}\label{eq:avg_fid}
    \bar{\mathcal{F}} = \frac{1}{N_\psi}\sum_{\psi} \mathcal{F}(\, \mathcal{E}\left[|\psi\rangle \langle \psi|\right]\,,\,U_{id}|\psi\rangle \langle \psi| U_{id}^\dagger\, ),
\end{equation}
where $U_{id}$ represents the unitary of the gate we are trying to implement and $\mathcal{E}[\cdot]$ is the quantum dynamical map of the noisy evolution corresponding to the chosen gates. The average should be taken over a set of states $\{|\psi\rangle\}$ which is close to the Haar random measure. 

The state fidelity is defined as 
\begin{equation}
    \mathcal{F}(\hat{\rho},\hat{\sigma}) = \left|\mathrm{tr}\sqrt{\sqrt{\hat{\rho}}\hat{\sigma} \sqrt{\hat{\rho}}}\right|^2,
\end{equation}
so that if one of the density matrices $\hat{\rho}$ or $\hat{\sigma}$ corresponds to a pure state, the state fidelity is the modulus squared of the overlap of the density matrix with the pure state.

However, the full Hilbert space of the hardware is much larger than the computational basis, meaning that it is not straightforward to reduce the whole system to its computational basis. Moreover, this is also relevant to the description of the measurement procedure in Eq. \ref{eq:measurement}.

In the case of a single qubit it is straightforward to define a qubitized density matrix in the computational subspace, by simply considering 
\begin{equation}\label{eq:qubitized_dm}
    [\hat{\rho}_\mathrm{comp}]_{ij} = \langle \phi_i  | \hat{\rho}|  \phi_j \rangle,
\end{equation}
where we consider the set of computational basis states for the single qubit case as $| \phi_i \rangle \in \{|0\rangle, |1\rangle \}$. Note that such a definition of the qubitized density matrix $\hat{\rho}_\mathrm{comp}$ has trace less than one, i.e. $\mathrm{tr}\left\{ \hat{\rho}_\mathrm{comp} \right\}\leq 1$, which is due to leakage outside of the computational subspace. We do not enforce the normalization when assessing the performance, meaning that the value of the fidelity is limited to the trace of the extracted computational subspace density matrix. Since the original matrix $\hat{\rho}$ is positive semi-definite, it follows that $\hat{\rho}_\mathrm{comp}$ is also positive semi-definite, as it is defined as a principal submatrix of $\hat{\rho}$ \cite{roger_johnson_matrix_analysis}. By this definition, the qubitized density matrix is also hermitian. If the normalization is also taken into account, such a qubitized density matrix is completely physical.

The above definition can be easily extended to the multi-qubit case with couplers by considering computational states where the coupler is not excited, namely $|\phi_i\rangle \in \{|\widetilde{000}\rangle,|\widetilde{001}\rangle, |\widetilde{100}\rangle, |\widetilde{101}\rangle \}$, with the ordering qubit 1, coupler, qubit 2. As before, the computational states are given by the eigenstates of the multi-qubit Hamiltonian and identified via the maximum overlap rule, which is why we have denoted these Hamiltonian eigenstates with a tilde.

However, there is also a different definition of the qubitized density matrix in this case, which is closer to the actual measured state, if one had access to it. Since we cannot probe the coupler states in any way, it would make sense to trace out the coupler degrees of freedom and then restrict the Hilbert space as in Eq. \ref{eq:qubitized_dm}, which would result in a slightly different definition. By this definition, for the two-qubit gate with one coupler, the qubitized density matrix is computed by
\begin{equation}\label{eq:qubitized_dm_trace}
    [\hat{\rho}_\mathrm{exp}]_{ij} = \langle \phi_i  | \mathrm{tr}_C\left\{\hat{\rho}\right\}|  \phi_j \rangle.
\end{equation}
The same arguments as in the previous definition still hold to show that this density matrix is physical. The full density matrix is first transformed into the eigenbasis of the full Hamiltonian and afterwards the trace over the coupler states is performed. The states $|\phi_i\rangle$ now no longer contain the coupler degrees of freedom, but are still defined in the Hamiltonian eigenbasis, and are therefore given by $|\phi_i\rangle \in \{|\widetilde{00}\rangle,|\widetilde{01}\rangle, |\widetilde{10}\rangle, |\widetilde{11}\rangle \}$.

While the latter definition from Eq. \ref{eq:qubitized_dm_trace} might be closer to the experimental measurement, we define the fidelity of an operation with the first definition from Eq. \ref{eq:qubitized_dm}, which accounts for the fact that any leakage into the coupler's excited states is undesired, since it may corrupt subsequent gate operations.

\section{Modelling many-body amplitude relaxation errors}\label{app:t1_model}

Since it is well-established that the main decoherence mechanism in transmon qubits is the coupling to a bath of two-level defects \cite{Muller_2019,cho2022,abdurakhimov_2022}, we consider that the transmon is coupled to a two-level system (TLS) bath, with a Hamiltonian of the form
\begin{equation}\label{eq:tls_bath_ham}
    \hat{H}_\mathrm{TLS} = \sum_{i \in \{Q_1,C,Q_2\}}\sum_{k} E_{ik} \sigma_{ik}^+ \sigma_{ik}^-.
\end{equation}
Here we are assuming that the TLSs are localized to each transmon, since they reside in the amorphous oxide layers of the various interfaces of each individual circuit element. Additionally, we are neglecting the phononic bath of the TLS, i.e. the intrinsic decoherence of the TLS themselves, and we omit the TLS which have an additional term proportional to $\sigma_{ik}^x$, as was done in Ref. \cite{abdurakhimov_2022}, and focus on the resonant TLS which contribute more to the decay. Note that the magnitude of the asymmetry in the standard tunnelling model is assumed to be distributed according to an $1/x$-type probability distribution, meaning that very small asymmetry energies are much more likely.

Since each TLS interacts with the transmon via electric fields, the interaction Hamiltonian of the system in the rotating wave approximation is given by
\begin{equation}\label{eq:spin_coupling_ham}
    \hat{H}_\mathrm{int} = \sum_{i \in \{Q_1,C,Q_2\}}\sum_{k} \chi_{ik} (\hat{a}_i^\dagger \hat{\sigma}^-_{ik} + \hat{a}_i\hat{\sigma}_{ik}^+).
\end{equation}
The form of the coupling in Eq. \ref{eq:spin_coupling_ham} is constructed to account for the observed excitation exchange between the transmon and a bath TLS \cite{Burnett_2019,abdurakhimov_2022}.

The coupling strength $\chi_{ik}$ was derived in \cite{abdurakhimov_2022,Muller_2019}
\begin{equation}
    \chi_{ik} = \frac{\delta_k {n_g}}{2}\sqrt{\omega_i E_{C_i}}.
\end{equation}
The constant $\delta_k n_g$ reflects the effect of the TLS electric dipole on the applied voltage on the superconducting island, while $\omega_i$ is the transmon frequency and $E_{C_i}$ the charging energy of the same transmon. While this means that the coupling constant depends on the frequency of the uncoupled transmon, the $\chi_{ik}$ is not a function of the excitation energy of each individual TLS $E_{ik}$.

We define the correlation function of the bath as
\begin{align}
    C_{ik\pm,i'k'\pm}(t) &= \nonumber \\ &\chi_{ik}\chi_{i'k'} \mathrm{tr}\left\{\, e^{-i \hat{H}_\mathrm{bath}t}\hat{\sigma}_{ik}^{\pm}e^{i \hat{H}_\mathrm{bath}t}\, \hat{\sigma}_{i'k'}^{\pm} \hat{\rho}_B \right\}.  
\end{align}
The bath density matrix is denoted by $\hat{\rho}_B$ and is assumed to be in a thermal state $\hat{\rho}_B \propto \exp{(-\beta \hat{H}_\mathrm{TLS})}$. Further assuming a diagonalized bath Hamiltonian of the form presented in Eq. \ref{eq:tls_bath_ham}, the correlation function is non-zero only if $i = i'$, $k = k'$ and for the operators preserving the mode occupation number, meaning that only the combinations $C_{ik+,ik-}$ and $C_{ik-,ik+}$ are possible. 

Focusing on the non-unitary dynamics, the rates $\Gamma^{ab}_1$ in Eq. \ref{eq:tqgT1_lndb} are computed as follows
\begin{equation}\label{eq:gamma_ab_def}
    \Gamma^{ab}_1 = \sum_{\substack{i \in \{Q_1,C,Q_2\} \\ p \in \{+,-\}}} \gamma_{ip}(\varepsilon_b - \varepsilon_a) 
    \langle a | \hat{a}_{i}^{-p} | b \rangle  \langle b | \hat{a}_{i}^p | a \rangle, 
\end{equation}
where the eigenenergies of the Hamiltonian from Eq. \ref{eq:tqg_ham} are denoted by $\hat{H}|a\rangle = \varepsilon_a|a\rangle$. The notation for the operators has been shortened to $\hat{a}^{-(+)} = \hat{a}^{(\dagger)} $ and we have defined the function

\begin{equation}
    \gamma_{ip}(\Delta\varepsilon) = \sum_k \mathrm{Re}\left\{ \int_0^\infty \mathrm{d}\tau \,C_{ikp,ik(-p)}(\tau)e^{i\,\Delta\varepsilon\, \tau}\right\} 
\end{equation}

It can also easily be verified that this definition of the rates $\Gamma^{ab}_1$ and jump operators in the uncoupled case when all $g_{ij} = 0$ reduces to the simpler local Lindblad equation from the single-qubit case in Eq. \ref{eq:T1_lndb}.

The function $\gamma_{ip}(\Delta\varepsilon)$ determines the decay rate of the eigenstate transition depending on its frequency, and can therefore be probed, in the uncoupled limit, by measuring the $T_1$ time at different frequencies of the qubit, as was done in Ref. \cite{carroll_2022}. In our TLS bath model, this function is given by 
\begin{equation}\label{eq:tls_bath_gamma}
    \gamma_{ip}(\Delta\varepsilon)  = 
    \begin{cases}
        \sum_k  \frac{\chi_{ik}^2}{1 + e^{-\beta E_{ik}}} \delta(E_{ik} - \Delta\varepsilon),& p= -,\\
        \sum_k \frac{\chi_{ik}^2}{1 + e^{\beta E_{ik}}} \delta(E_{ik}+\Delta\varepsilon),& p = +,
    \end{cases}
\end{equation}
where we have used the delta function $\delta(\cdot)$ to simplify the notation, and as an approximation, we consider the continuous bath approximation from now on, so that $\sum_{ik} \rightarrow \int \mathrm{d}E_{i} \mathrm{d}(\delta_i n_g) P(E_{i},\delta_i n_g)$, where $P(E_i)$ is the continuous probability distribution of the TLS energy, which is typically assumed to be uniform \cite{cho2022}, and independent of the coupling parameter $\delta n_g$ in the standard tunnelling model, i.e. $ P(E_{i},\delta_i n_g) = P(E_i)P(\delta_i n_g)$. In reality, the continuous approximation is not needed, since the resonance with an individual TLS described by the delta function in Eq. \ref{eq:tls_bath_gamma} is broadened due to the intrinsic decoherence of the TLS \cite{shnirman_2005}. Additionally, since the parameters of the TLS environments are random \cite{cho2022}, a complete description of the TLS bath would imply knowing the parameters of hundreds of TLSs. 

While we have presented a form of the transmon-TLS coupling in Eq. \ref{eq:spin_coupling_ham}, we still need to connect the above model to an experimental observation. This is done by assuming that the system is completely uncoupled when the $T_1$ time is measured, which results in $\langle \chi_{i}^2\rangle = \Gamma_1^i$.
with the constant $\Gamma_1^i = 1/T_1^i $ of the corresponding transmon $i \in \{Q_1,C,Q_2\}$ measured in the uncoupled regime, where the hybridization between the states can be neglected. We have defined the averaged coupling strength as $\langle \chi_{i}^2\rangle = \int \mathrm{d}(\delta_i n_g)\, \chi_{i}^2(\delta_i n_g) P(\delta_i n_g)$, and therefore the actual distribution of the coupling strengths of the TLSs is irrelevant, as long as the coupling strength is independent of the TLS energy $E_{ik}$. 

The general form of the rates in Eq. \ref{eq:tqgT1_lndb}, obtained by using the above result in Eq. \ref{eq:gamma_ab_def} is
\begin{align}
    \Gamma^{ab}_1 &\approx \sum_{i \in \{Q_1,C,Q_2\}} \frac{\Gamma_1^i}{1 + e^{-\beta (\varepsilon_b - \varepsilon_a)}}
    |\langle a | \hat{a}_{i} | b \rangle |^2 \Theta(\varepsilon_b - \varepsilon_a) \nonumber \\
    & +  \frac{\Gamma_1^i}{1 + e^{\beta (\varepsilon_a - \varepsilon_b)}} |\langle a | \hat{a}_{i}^\dagger | b \rangle |^2 \Theta(\varepsilon_a - \varepsilon_b).
\end{align}
The above equation therefore approximately models the decay rates between different eigenstate transitions during the operation of a two-qubit gate. The many-body effects are captured by the matrix elements of the operators $\hat{a}^{(\dagger)}$, and the transition energy dependence is a result of the physical TLS model. The exact nature of the bath (spins, bosonic or fermionic) becomes relevant in the regime where the transition energy becomes comparable to the temperature $|\varepsilon_a - \varepsilon_b| \sim 1/\beta $. In our simulations, this is a common occurrence for most transitions within a fixed excitation subspace in the near-resonant regime during the operation of the gate, as the frequency corresponding to a typical cryostat temperature of 20 mK is approximately 0.5 GHz.

\section{Details of the Experimental Implementation}\label{app:exp_details}

Here we present the details of the experimental implementation of the protocol presented in Sec.~\ref{sec:budgeting_demo}. While the data for both single- and two-qubit gates was obtained from the same device, both sets were not collected at the same time, meaning that the parameters of the device might have changed between the two measurements.

\subsection{Simulation Parameters}

\subsubsection{Single-Qubit Gates}\label{app:single_qubit_gate_details}
Here we list the distributions of the Hamiltonian, error and SPAM error parameters used to obtain the simulation data for the single-qubit gate results in Fig.~\ref{fig:experiments}. The parameters were chosen to be close to the ones of current superconducting hardware, while also allowing for a large variety of gate implementations. More information about each of the parameters listed here can be found in Appendix~\ref{app:sqg_noise_model}.

Firstly, we generate a set of Hamiltonian parameters according to Table~\ref{tab:sqg_ham_params}. While the $\sigma$ of the Gaussian DRAG pulse is fixed to $8.8$~ns, we then optimize the I component amplitude of the pulse $A$ as well as the Q component amplitude $\beta$. This is done to mimic the calibration of the gate where only the optimized parameters are varied.

\begin{table}[h]
\caption{\label{tab:sqg_ham_params}
The ranges and distributions of the Hamiltonian parameters for the single-qubit gate simulations. All parameters are sampled independently from each other. The distributions are either uniform $\mathcal{U}(a,b)$ from the interval $(a,b)$ or normal $\mathcal{N}(\mu,\sigma)$, with the parameters of the distribution given in the second column. The units of each parameter are given in the last column. The 'or' statement is used to indicate that the parameters are sampled from one of the two distributions depending on the qubit being characterized.
}
\begin{ruledtabular}
\begin{tabular}{ccc}
Parameter & Distribution & Units \\
\hline\hline
$\omega$ & $\mathcal{N}(4.1,0.02)$ or $\mathcal{N}(4.3,0.02)$ & GHz \\
$\alpha$ & $\mathcal{N}(-0.213,0.002)$ or $\mathcal{N}(-0.216,0.002)$ & GHz \\
\end{tabular}
\end{ruledtabular}
\end{table}
The frequencies and their variability of the single-qubit Hamiltonian in Table~\ref{tab:sqg_ham_params} were chosen to match the historically observed magnitude of the fluctuations of the device in the previous weeks before the experiments were performed. The variability of the anharmonicity is also low due to the increased accuracy with which this parameter can be controlled in fabrication.

\begin{table}[h]
\caption{\label{tab:sqg_error_params}
The ranges and distributions of the error parameters for the single-qubit gate simulations. All parameters are sampled independently from each other. The distributions are either uniform $\mathcal{U}(a,b)$ from the interval $(a,b)$ or normal $\mathcal{N}(\mu,\sigma)$, with the parameters of the distribution given in the second column. The units of each parameter are given in the last column. 
}
\begin{ruledtabular}
\begin{tabular}{ccc}
Parameter & Distribution & Units \\
\hline\hline
$T_1$ & $\mathcal{N}(35,5.3)$ & $\mu$s \\
$T_\phi$ & $\mathcal{N}(50,8)$ & $\mu$s \\
$T_{1/f}$ & $\mathcal{N}(11,1.4)$ & $\mu$s \\

$\epsilon_A$ & $\mathcal{N}(0,3.5)$ & deg \\
$\epsilon_\beta$ & $\mathcal{N}(0,0.03)$ & / \\

$\epsilon_\mathrm{phase}$ & $\mathcal{N}(0,0.5)$ & deg \\

$\delta\omega$ & $\mathcal{N}(0,20)$ & kHz 
\end{tabular}
\end{ruledtabular}
\end{table}
Historical data from both characterized qubits was used for the variability of the coherence times in Table~\ref{tab:sqg_error_params}, while the magnitude of the coherent errors was based on the assumption that these are the main contribution to the infidelity in conjunction with the historical gate fidelity. An ensemble of 1000 two-level fluctuators (TLFs) was used, with the lowest frequencies determined by 10$\times$ the duration of the experiment and the highest frequency being 10 GHz, since each gate was simulated in 1 ns steps, as described in Appendix~\ref{app:simulations}. The frequency ranges are chosen such that the generated dynamics follow the $1/f$ spectrum in the relevant frequency ranges. The coupling strength of all the fluctuators was fixed and determined by the specified $T_{1/f}$ coherence time. Altogether, 500 trajectories were sampled to observe convergence.

\begin{table}[h]
\caption{\label{tab:sqg_spam_params}
The ranges and distributions of the SPAM error parameters for the single-qubit gate simulations. All parameters are sampled independently from each other. The distributions are either uniform $\mathcal{U}(a,b)$ from the interval $(a,b)$ or normal $\mathcal{N}(\mu,\sigma)$, with the parameters of the distribution given in the second column. The units of each parameter are given in the last column. 
}
\begin{ruledtabular}
\begin{tabular}{ccc}
Parameter & Distribution & Units \\
\hline\hline
$T_\mathrm{eff}$ & $\mathcal{N}(40,2.5)$ & mK \\
$P(0|1)$ & / & / \\
$P(1|0)$ & / & / \\
\end{tabular}
\end{ruledtabular}
\end{table}
Regarding the SPAM errors listed in Table~\ref{tab:sqg_spam_params}, the effective temperature $T_\mathrm{eff}$ of the qubits and coupler was chosen based on measurements of similar devices. The resulting initial excited state populations of the qubits were in the range of $0.5\%$ to $1\%$. The measurement errors were not included in the single-qubit simulations as the gate errors were amplified accordingly, and high-accuracy was achieved in the results displayed in Fig.~\ref{fig:experiments} even without including this effect.

The training set for the GPR model contained a total of 450 simulated gates.

\subsubsection{Two-Qubit Gates}
Here we list the distributions of the Hamiltonian, error and SPAM error parameters used to obtain the simulation data for the two-qubit gate results in Fig.~\ref{fig:experiments}. More information about each of the parameters listed here can be found in Appendix~\ref{app:tqg_noise_model}. 

The parameters were chosen to be close to the ones of current superconducting hardware, while also allowing for a large variety of gate implementations. First a set of random Hamiltonian parameters is drawn, and the system is diagonalized to find the coupler idling frequency $\omega_C^\mathrm{idle}$. If the coupler idling frequency was higher than the generated maximal coupler frequency, the process was repeated. After a set of random error parameters is generated, the time $\tau_c$ and pulse amplitude $A$ were optimized to maximize the gate fidelity, and small deviations from the ideal pulse parameters were added afterwards. If the resulting two-qubit gate fidelity was smaller than 95\%, the whole process was repeated until a set of parameters resulting in a gate fidelity above 95\% was obtained. The starting guess for the pulse parameters was always a total pulse duration of $\tau = \tau_c + 2\tau_b = 40$~ns with a rise time $\tau_b = 10$~ns. When the experimental data was gathered, the two-qubit gate duration was fixed to 80~ns. The simulations reflect this: if the total duration of the optimized pulse was smaller than 80~ns, the qubits were left idling for the remaining time, i.e. the pulse was padded with a zero-amplitude segment up to the full 80~ns gate duration. Conversely, if the total pulse duration after optimization exceeded 80~ns, the parameter set was discarded, since such a gate would not fit within the fixed gate duration.

\begin{table}[h]
\caption{\label{tab:tqg_ham_params}
The ranges and distributions of the Hamiltonian parameters from Eq.~\ref{eq:tqg_ham} for the two-qubit gate simulations. All parameters are sampled independently from each other. The distributions are either uniform $\mathcal{U}(a,b)$ from the interval $(a,b)$ or normal $\mathcal{N}(\mu,\sigma)$, with the parameters of the distribution given in the second column. The units of each parameter are given in the last column. 
}
\begin{ruledtabular}
\begin{tabular}{ccc}
Parameter & Distribution & Units \\
\hline\hline
$\omega_{Q_1}$ & $\mathcal{N}(4.1,0.01)$ & GHz \\
$\omega_{C}^\mathrm{max}$ & $\mathcal{N}(6.4,0.3)$ & GHz \\
$\omega_{Q_2}$ & $\mathcal{N}(4.3,0.01)$ & GHz \\

$\alpha_{Q_1}$ & $\mathcal{N}(-0.216, 10^{-3})$ & GHz \\
$\alpha_{C}$ & $\mathcal{N}(-0.108, 10^{-3})$ & GHz \\
$\alpha_{Q_2}$ & $\mathcal{N}(-0.213, 10^{-3})$ & GHz \\

$\beta_{Q_1C}$ & $\mathcal{N}(2 \cdot 10^{-2}, 10^{-3})$ & / \\
$\beta_{Q_2 C}$ & $\mathcal{N}(2.1 \cdot 10^{-2}, 10^{-3})$ & / \\     
$\beta_{Q_1 Q_2}$ & $\mathcal{N}(2 \cdot 10^{-3}, 10^{-4})$ & / \\       
\end{tabular}
\end{ruledtabular}

\end{table}
Looking at Table~\ref{tab:tqg_ham_params}, the variability of the anharmonicities is relatively low as the transmon anharmonicity depends mostly on the charging energy and therefore the shape of the capacitor pads \cite{Koch_2007}, which can be accurately controlled during fabrication. On the other hand, the coupling parameters $\beta$ are much less certain as the cross-capacitances between the circuit elements are more difficult to control. Since a diabatic CZ gate can only be implemented in the regime where $|\omega_{Q_1} - \omega_{Q_2}| \sim -\alpha_{Q_2}$, the generated qubit frequency difference is therefore fixed by the anharmonicity. The low variability of the qubit frequencies is based on the historical data of the frequencies from the device. On the other hand, the coupler maximal frequency is more variable due to the variability of the Josephson energy during fabrication, and the fact that this frequency is not routinely measured as part of the calibration procedure.

\begin{table}[h]
\caption{\label{tab:tqg_error_params}
The ranges and distributions of the error parameters for the two-qubit gate simulations. All parameters are sampled independently from each other. The distributions are either uniform $\mathcal{U}(a,b)$ from the interval $(a,b)$ or normal $\mathcal{N}(\mu,\sigma)$, with the parameters of the distribution given in the second column. The units of each parameter are given in the last column. In the unlikely case that a generated parameter is unphysical (e.g. a negative decay time), it is resampled until a positive value is obtained.
}
\begin{ruledtabular}
\begin{tabular}{ccc}
Parameter & Distribution & Units \\
\hline\hline
Qubit 1 $T_1$ & $\mathcal{U}(35,44)$ & $\mu$s \\
Coupler $T_1$ & $\mathcal{U}(10,30)$ & $\mu$s \\
Qubit 2 $T_1$ & $\mathcal{U}(22,30)$ & $\mu$s \\

Qubit 1 $T_\phi$ & $\mathcal{N}(43,12)$ & $\mu$s \\
Coupler $T_\phi$ & $\mathcal{U}(10,30)$ & $\mu$s \\
Qubit 2 $T_\phi$ & $\mathcal{N}(15,3)$ & $\mu$s \\

Qubit 1 $T_{1/f}$ & $\mathcal{N}(14,6)$ & $\mu$s \\
Coupler $T_{1/f}$ & $\mathcal{U}(0.6,10)$ & $\mu$s \\
Qubit 2 $T_{1/f}$ & $\mathcal{N}(6.7,2)$ & $\mu$s \\

Qubit 1 Virtual Z Error & $\mathcal{N}(0,2)$ & deg \\
Qubit 2 Virtual Z Error & $\mathcal{N}(0,2)$ & deg \\

Pulse amplitude error $\delta A$ & $\mathcal{N}(0,0.005)$ & GHz \\
Pulse duration error $\delta \tau$ & $\mathcal{N}(0,0.02)$ & ns \\

Flux tail $A_n$ & $\mathcal{N}(0,0.005)$ & / \\
Flux tail $\tau_n$ & $\mathcal{U}(50,200)$ & ns \\

\end{tabular}
\end{ruledtabular}

\end{table}
Table~\ref{tab:tqg_error_params} contains the parameters of the error sources used in the simulation. The parameters which are routinely measured generally have a lower variability and the distribution was chosen to mimic the device performance in the previous 3 weeks before the experimental data was gathered. On the other hand, for the parameters which are not measured, their maximal contribution was inferred based on the worst-case scenario that the observed gate infidelities are a result of that error exclusively.

The variability and range of the computational transmons coherence times (Markovian $T_1$ and $T_\phi$, as well as $1/f$- noise induced dephasing time $T_{1/f}$) are based on the historical data of the device for the previous 3 weeks. On the other hand, as the coupler coherence is not routinely measured, a larger variability is assumed for the coupler coherence times. The TLF ensemble was generated identically as in the single-qubit case described in Appendix~\ref{app:single_qubit_gate_details}. The magnitude of the virtual Z errors was chosen such that this error would, in the worst case, correspond to the maximal observed infidelity of the gate historically, as no measurements of this error were available. The same logic was applied to the deviation of the optimal pulse amplitude $\delta A$, while the pulse duration error $\delta \tau$ was determined by the sampling rate of the electronics. The parameters of the flux tail were chosen such that the resulting error would be on the order of the observed infidelity of the gate, while also allowing for a large variability in the parameters. We note that only one exponential contribution to the flux tail was considered.

\begin{table}[h]
\caption{\label{tab:tqg_spam_error_params}
The ranges and distributions of the SPAM error parameters for the two-qubit gate simulations. All parameters are sampled independently from each other. The distributions are either uniform $\mathcal{U}(a,b)$ from the interval $(a,b)$ or normal $\mathcal{N}(\mu,\sigma)$, with the parameters of the distribution given in the second column. The units of each parameter are given in the last column. The definitions of the parameters can be found in Appendix~\ref{app:state_preparation} and Appendix~\ref{app:measurement}.
}
\begin{ruledtabular}
\begin{tabular}{ccc}
Parameter & Distribution & Units \\
\hline\hline
Qubit 1 $T_\mathrm{eff}$ & $\mathcal{N}(40,2)$ & mK \\
Coupler $T_\mathrm{eff}$ & $\mathcal{N}(40,2)$ & mK \\
Qubit 2 $T_\mathrm{eff}$ & $\mathcal{N}(40,2)$ & mK \\
\hline
Qubit 1 $P(0|1)$ & $\mathcal{N}(1.6,0.16)$ & \% \\
Qubit 2 $P(0|1)$ & $\mathcal{N}(3,0.3)$ & \% \\
Qubit 1 $P(1|0)$ & $\mathcal{N}(0.6,0.06)$ & \% \\
Qubit 2 $P(1|0)$ & $\mathcal{N}(0.4,0.04)$ & \% \\
\end{tabular}
\end{ruledtabular}

\end{table}
The SPAM error parameters are listed in Table~\ref{tab:tqg_spam_error_params}. The effective temperature $T_\mathrm{eff}$ of the qubits and coupler was chosen based on measurements of similar devices. The resulting initial excited state populations of the qubits were in the range of $0.5\%$ to $1\%$. The measurement error parameters were chosen based on the historical data of the device for the previous 3 weeks before the experimental data was gathered.

All together 720 gates were simulated and included in the training set.

\subsection{Data Pre-Processing}
Before training the GPR model, the simulated data was pre-processed according to the following steps:
\begin{enumerate}
    \item A standard scaler was applied to the simulated characterization results to normalize the inputs to a zero-mean and unit-variance distribution.
    \item A standard scaler in conjunction with a power-$1/2$ transformer was applied to the simulated gate infidelities.
    \item For the two-qubit case, the tomographical experiments were used to reconstruct the qubitized density matrix after each experiment.
    \item For the two-qubit case, a correlation analysis was performed based on the Chatterjee correlation coefficient \cite{chatterjee2020}, which is a non-parametric correlation coefficient that can capture non-linear relationships between the features and the target variable. The features with the lowest correlation were removed from the dataset to reduce the dimensionality of the input space and improve the performance of the GPR model. The 20 most correlated inputs were retained. This was not required for the single-qubit case, as only 26 experiments were initially considered.
    \item Outliers were removed from the dataset if the z-score was larger than 3. This was done to improve the performance of the GPR model, as outliers can have a significant impact on the training process and lead to overfitting.
\end{enumerate}

\subsection{GPR Implementation}

We employ different kernels for the GPR model, as described in Eq.~\ref{eq:kernel}. Besides the RBF kernel listed in Eq.~\ref{eq:rbf_kernel}, we also use the following kernel functions:

\begin{align}
    k_\mathrm{WhiteKernel}&(\mathbf{x}_i,\mathbf{x}_j) = \sigma^2 \delta(\mathbf{x}_i,\mathbf{x}_j) \\
    k_\mathrm{ExpSineSquared}&(\mathbf{x}_i,\mathbf{x}_j) = \nonumber \\
    & \sigma^2 \exp\left(-\frac{2\sin^2(\pi |\mathbf{x}_i - \mathbf{x}_j|/p)}{l^2}\right). 
\end{align}

Specifically, we use a WhiteKernel to account for the noise in the data, and an ExpSineSquared kernel to capture the periodicity in the data, which can arise as a result of coherent errors. For the $\sqrt{X}$ gate results in Fig.~\ref{fig:experiments}, the kernel is given by RBF + RBF + RBF + WhiteKernel. The choice of kernel for the CZ results was determined by trial and error by optimizing the performance on simulated data. The exact kernel constructions used for the results in Fig.~\ref{fig:experiments} are given in Table~\ref{tab:tqg_kernels} for the two-qubit case. The optimizer was restarted 40 and 100 times for the single-qubit and two-qubit cases, respectively, to find the optimal hyperparameters of the kernel.

\begin{table}
\caption{\label{tab:tqg_kernels}
The kernel constructions used for the CZ characterization results in Fig.~\ref{fig:experiments}. }
\begin{ruledtabular}
\begin{tabular}{cc}
Error Source & Kernel  \\
\hline\hline
Qubit 1 $T_1$ & RBF + RBF + RBF \\
Coupler $T_1$ & RBF + RBF + RBF \\
Qubit 2 $T_1$ & RBF + RBF + RBF \\

Qubit 1 $T_\phi$ & RBF + RBF + RBF \\
Coupler $T_\phi$ & RBF + RBF + RBF \\
Qubit 2 $T_\phi$ & RBF + RBF + RBF \\

Qubit 1 $T_{1/f}$ & RBF + RBF + RBF \\
Coupler $T_{1/f}$ & RBF + RBF + RBF \\
Qubit 2 $T_{1/f}$ & RBF + RBF + RBF \\

Qubit 1 Virtual Z & RBF + WhiteKernel + ExpSineSquared \\
Qubit 2 Virtual Z & RBF + WhiteKernel + ExpSineSquared \\

Flux tails $A_n$ & RBF + RBF + WhiteKernel \\

Control & RBF + WhiteKernel + ExpSineSquared \\

\end{tabular}
\end{ruledtabular}

\end{table}

\bibliography{biblio}

\end{document}